\pdfoutput=1
\documentclass[11pt]{article}

\usepackage[final]{acl}

\usepackage{times}
\usepackage{latexsym}
\usepackage{float}
\usepackage{booktabs}   % For \toprule, \midrule, \bottomrule
\usepackage{siunitx}    % For the S column type (numeric alignment)
\usepackage{float}      % For the [H] placement option
\usepackage{tabularx}   % For the X column type (text wrapping, auto-width)
\usepackage{caption}
\usepackage{multirow}
\usepackage{multicol}
\usepackage{amsmath}
\usepackage{amsfonts}
\usepackage{amssymb}
\usepackage{makecell}
\usepackage{xcolor}
\usepackage{colortbl}
\usepackage{arydshln}
\usepackage{stfloats}
\usepackage{subcaption}
\usepackage{enumitem}
\usepackage{makecell} % For line breaks in cells
\usepackage{pbox}     % Alternative wrapping method
\usepackage{longtable} % Optional for multi-page table
\definecolor{bestwercolor}{HTML}{778899}
\definecolor{bestsimcolor}{HTML}{BDB0D0}

\usepackage{array} % For custom column types
\usepackage{siunitx} % For aligning numbers by decimal point and handling ±

\newcolumntype{P}{S[table-format=-1.2, table-space-text-post=\si{\pm}0.22]} % Adjust table-space-text-post if needed for wider error margins

\usepackage[T1]{fontenc}
\usepackage[utf8]{inputenc}

\usepackage{microtype}

\usepackage{inconsolata}
\usepackage{booktabs}

\usepackage{graphicx}

\title{VoiceCraft-X: Unifying Multilingual, Voice-Cloning Speech Synthesis \\ and Speech Editing}
\author{
\textbf{Zhisheng Zheng}\textsuperscript{1},
\textbf{Puyuan Peng}\textsuperscript{1},
\textbf{Anuj Diwan}\textsuperscript{1},
\textbf{Cong Phuoc Huynh}\textsuperscript{2},\\
\textbf{Xiaohang Sun}\textsuperscript{2},
\textbf{Zhu Liu}\textsuperscript{2}, 
\textbf{Vimal Bhat}\textsuperscript{2}, 
\textbf{David Harwath}\textsuperscript{1}\thanks{Corresponding author}
\\
\\
 \textsuperscript{1}University of Texas at Austin, 
 \textsuperscript{2}Amazon\\ 
}

\makeatletter
\def\@fnsymbol#1{\ensuremath{\ifcase#1\or \dagger\or *\or \ddagger\or \S \or \P \or \Vert \or **\or \dagger\dagger \or \ddagger\ddagger\fi}}
\makeatother

\begin{document}
\maketitle
\begin{abstract}
We introduce VoiceCraft-X, an autoregressive neural codec language model which unifies multilingual speech editing and zero-shot Text-to-Speech (TTS) synthesis across 11 languages: English, Mandarin, Korean, Japanese, Spanish, French, German, Dutch, Italian, Portuguese, and Polish. VoiceCraft-X utilizes the Qwen3 large language model for phoneme-free cross-lingual text processing and a novel token reordering mechanism with time-aligned text and speech tokens to handle both tasks as a single sequence generation problem. The model generates high-quality, natural-sounding speech, seamlessly creating new audio or editing existing recordings within one framework. VoiceCraft-X shows robust performance in diverse linguistic settings, even with limited per-language data, underscoring the power of unified autoregressive approaches for advancing complex, real-world multilingual speech applications. Audio samples are available at \url{https://zhishengzheng.com/voicecraft-x/}.

\end{abstract}

\section{Introduction}
Highly realistic speech generation is an indispensable technology for voice assistants, content dubbing, accessibility tools, and creative media. Speech generation can be broken down into several sub-problems: \emph{creating} new audio via Text-To-Speech synthesis (TTS) or \emph{editing}  part of an existing recording while ensuring voice consistency with the remainder of the original speech. Despite their shared goal of producing natural speech, TTS and speech editing are typically treated as \emph{separate} problems, especially in multilingual settings, which leaves practitioners without a \emph{single} model that can both edit and synthesize speech across languages.

Over the past several years, the quality of TTS models has improved significantly, particularly in the zero-shot setting in which a model generates speech in a new speaker's voice given a short (e.g. 3 second) audio prompt. Transformer-based neural networks have been central to this progress, leading to three broad paradigms: (i) autoregressive (AR), (ii) non-autoregressive (Non-AR), and (iii) hybrid models.  
AR models, such as VALL-E~\citep{wang2023neural} and its successors~\citep{zhang2023speak, han2024vall, xin2024rall, chen2024vall, song2025ella, yang2025pseudo}, generate frame-level speech tokens sequentially, where the tokens are typically derived from a neural audio codec~\citep{defossez2022high, zeghidour2021soundstream, zhang2023speechtokenizer}. These models are able to perform voice-cloning TTS via Transformer language models' in-context learning ability, demonstrating high-quality speech synthesis. Non-AR models include flow-matching models such as F5-TTS~\citep{chen2024f5}, as well as diffusion models such as NaturalSpeech 2/3~\citep{shen2023naturalspeech, ju2024naturalspeech}. These models predict all tokens representing an utterance in parallel via iterative refinement. Hybrid approaches such as Seed-TTS~\citep{anastassiou2024seed}, CosyVoice~\citep{du2024cosyvoice1, du2024cosyvoice2} and MaskGCT~\citep{wang2024maskgct} aim to combine the strengths of both paradigms.  
While these models deliver impressive zero-shot quality, most of the models are either monolingual or focus on a handful of high-resource languages such as English and Chinese. This is likely due to the fact that these models are data-hungry, often requiring \text{10K-100K} hours of training speech for SOTA performance.

The quest for broader linguistic inclusivity across the world's 7,000 spoken languages~\citep{Eberhard2024} has driven research in multilingual speech generation. Efforts include curating large corpora (e.g., VoxPopuliTTS~\citep{liu2025voxpopulitts}, Fish-Speech~\citep{liao2024fish}) and training multilingual TTS architectures like VoiceBox~\citep{le2023voicebox}, CLAM-TTS~\citep{kim2024clam} and XTTS~\citep{casanova2024xtts}.  
Yet even the most capable multilingual systems treat \emph{speech editing} as a separate task—or ignore it altogether—leaving users without a unified solution.

In this paper we address this gap, by introducing \textbf{VoiceCraft-X}, a unified autoregressive neural codec language model that performs \emph{both} speech editing and zero-shot TTS in \textbf{11 languages}: English (en), Mandarin (zh), Korean (ko), Japanese (ja), Spanish (es), French (fr), German (de), Dutch (nl), Italian (it), Portuguese (pt) and Polish (pl).  
Our contributions are threefold:  
\begin{enumerate}[itemsep=-1.5pt, topsep=0pt]
    \item We introduce VoiceCraft-X, a single autoregressive model that unifies multilingual speech editing and zero-shot Text-to-Speech (TTS) across 11 languages.
    \item Our approach leverages the Qwen3 large language model for cross-lingual text processing, without the need for phonetic pronunciation lexicons. We also propose a novel token reordering mechanism that time-aligns text and speech, enabling a unified sequence generation approach for both editing and synthesis.
    \item We demonstrate VoiceCraft-X's robust generation of high-quality, natural-sounding speech across diverse languages, even with limited per-language data, and will release our code and model to the community.
\end{enumerate}

\section{Related Work}
\subsection{Speech Editing}
Speech editing aims to correct mispronunciations, stutters, or recording artifacts while producing speech that is indistinguishable from natural audio. Recent approaches leverage Transformer and diffusion architectures. \citet{borsos2022speechpainter} perform audio infilling with a Transformer that maintains speaker identity and prosody, generalizing to unseen speakers. ~\citet{le2023voicebox} use flow matching for versatile speech infilling, and \citet{peng2024voicecraft} show that a neural-codec language model with token infilling can concurrently handle editing and synthesis. F5-TTS~\citep{chen2024f5} and MaskGCT~\citep{wang2024maskgct} extend this idea with flow-matching or diffusion, respectively. Despite these advances, most works are monolingual, motivating a unified multilingual solution.

\subsection{Zero-Shot Speech Synthesis}
The zero-shot Text-to-Speech (TTS) synthesis task entails generating speech in a new speaker's voice from a short audio prompt, without assuming that the new speaker was seen during training. Recent progress is largely driven by Transformer-based neural networks, falling into autoregressive (AR), non-autoregressive (non-AR), and hybrid.

Autoregressive (AR) models generate speech tokens sequentially. VALL-E~\citep{wang2023neural} pioneered neural codec language models for high-quality zero-shot TTS via in-context learning, with subsequent works~\citep{zhang2023speak, han2024vall, chen2024vall, xin2024rall, song2025ella, kharitonov2023speak, lajszczak2024base, peng2024voicecraft, guo2024fireredtts} further refining this paradigm. Non-Autoregressive (Non-AR) models aim for faster generation by predicting tokens in parallel or using iterative refinement. Examples include flow-matching models like VoiceBox~\citep{le2023voicebox} and diffusion-based models such as NaturalSpeech 2~\citep{shen2023naturalspeech}, NaturalSpeech 3~\citep{ju2024naturalspeech}, and DiTTo-TTS~\citep{lee2024ditto}. Other notable non-AR approaches include Unicats~\citep{du2024unicats}, SimpleSpeech~\citep{yang2024simplespeech1, yang2024simplespeech2}, E2-TTS~\citep{eskimez2024e2}, F5-TTS~\citep{chen2024f5} and Mega-TTS 3~\citep{jiang2025megatts}. Hybrid systems combine aspects of both AR and non-AR methods. Seed-TTS~\citep{anastassiou2024seed} uses a two-stage architecture, while CosyVoice~\citep{du2024cosyvoice1, du2024cosyvoice2} and MaskGCT~\citep{wang2024maskgct} also represent efforts to balance quality, speed, and controllability. In this work, VoiceCraft-X follows the codec language modeling method of VoiceCraft~\citep{peng2024voicecraft} and enables high-quality, zero-shot multilingual speech synthesis within its unified editing and generation framework.

\subsection{Multilingual Speech Generation}
Prior work on multilingual speech synthesis largely pursues two complementary goals: (i) expanding language coverage and (ii) achieving zero-shot robustness to unseen speakers and languages.

On the data side, \citet{saeki2024extending} show that pairing self-supervised speech representations with unsupervised text alignment scales TTS to 100\,+ languages, even when only scant transcriptions exist. Large curated corpora amplify these gains: VoxPopuliTTS~\citep{liu2025voxpopulitts} refines 30,000 hours of English, French and Spanish speech; Fish-Speech ~\citep{liao2024fish} goes further, training on 720,000 hours while using an LLM to sidestep language-specific G2P rules. Model architectures have evolved in parallel. VoiceBox~\citep{le2023voicebox} adopts non-autoregressive flow matching, delivering cross-lingual zero-shot TTS in six languages via in-context learning. XTTS~\citep{casanova2024xtts}, extending Tortoise~\citep{betker2023better}, combines a Perceiver Resampler with a speaker-consistency loss to reach 16 languages with speaker cloning. CLAM-TTS~\citep{kim2024clam} improves codec language model compression with probabilistic residual vector quantization, enabling single-step multi-token generation. However, these models often treat synthesis as a distinct task from speech editing. The challenge of \emph{unifying} high-quality, multilingual speech editing with robust multilingual speech synthesis within a single, open-source, and fully autoregressive model architecture remains largely unaddressed.

\section{Method}
\subsection{Overview}
\begin{figure*}[htp]
\vspace{-0.3cm}
    \centering
    \includegraphics[width=0.95\linewidth]{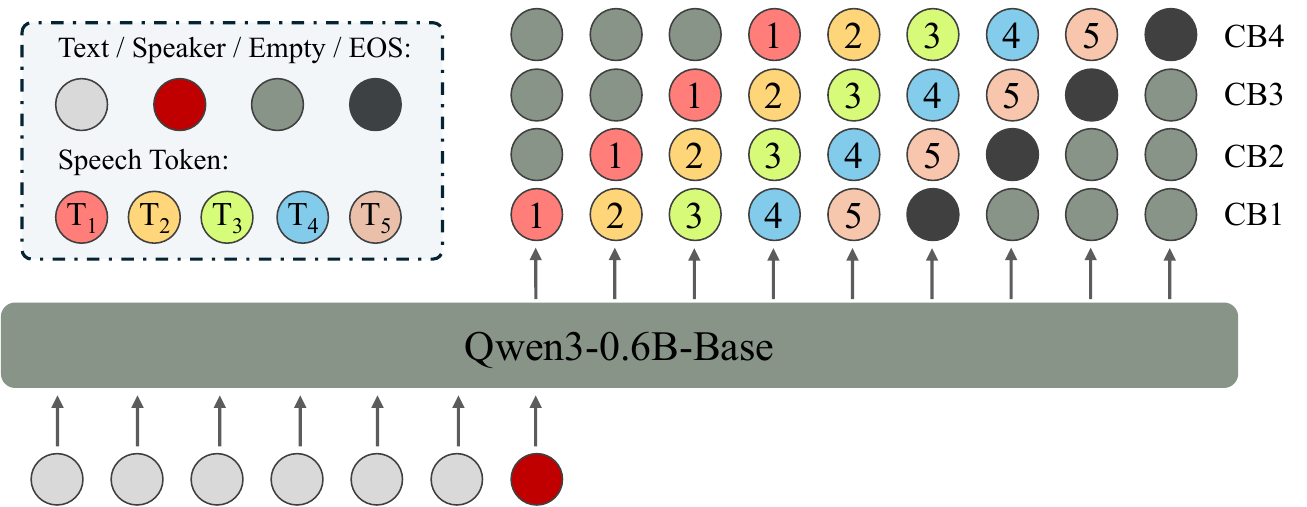}
    \caption{\textbf{Architecture Overview.} This diagram illustrates the training process for the VoiceCraft-X model. The model takes text and a speaker embedding as input and is trained to predict sequences of speech tokens. The labels CB1-CB4 represent codec tokens from different codebooks.}
    \label{fig:architecture}
    \vspace{-0.5cm}
\end{figure*}

VoiceCraft-X evolves VoiceCraft~\citep{peng2024voicecraft} into a truly multilingual speech-editing and synthesis system, treating both tasks as a single sequence-generation problem over neural codec tokens. The core of this system, as illustrated in Figure~\ref{fig:architecture}, is the Qwen3~\citep{qwen3} large language model. Qwen3 natively supports text input in $119$ languages and dialects, which we leverage as the cross-lingual input text tokenizer for VoiceCraft-X. This eliminates the cumbersome phoneme-conversion step that was integral to the original VoiceCraft, resulting in a simplified pipeline with a shared tokenizer across languages, without the need to curate pronunciation lexicons for each language.

A further key innovation in VoiceCraft-X is its enhanced data layout: it interleaves text tokens and speech tokens in a single, time-ordered stream, whereas VoiceCraft reordered only the speech tokens. Enforcing this alignment between linguistic content and its acoustic realization yields more consistent and natural-sounding speech.

\subsection{Speaker Embedding}
In addition to the speech tokens representing the prompt speech, VoiceCraft-X also takes as input a speaker embedding vector extracted from this prompt speech. We follow the approach of CosyVoice~\citep{du2024cosyvoice1} by using a pre-trained voiceprint model\footnote{\url{https://www.modelscope.cn/models/iic/CosyVoice-300M/file/view/master/campplus.onnx}} to extract the speaker embedding. The resulting vector is then passed through a linear projection layer. This projection maps the speaker embedding to match Qwen3's input dimension.

\subsection{Speech Tokenization}
%\david{Say a few sentences here about how we tokenize the speech}
%The first step for both training and inference with VoiceCraft-X is to tokenize the input utterance. 
We utilize the EnCodec~\citep{defossez2022high} neural audio codec model to tokenize the input utterance. Specifically, we train a modified version of the tokenizer which outputs a sequence of four parallel token streams at a 50Hz framerate. The tokens are discretized with residual vector quantization (RVQ) with a vocabulary size of 2048 at each quantization layer.

\subsection{Token Reordering}
\begin{figure*}[htp]
\vspace{-0.3cm}
    \centering
    \includegraphics[width=1\linewidth]{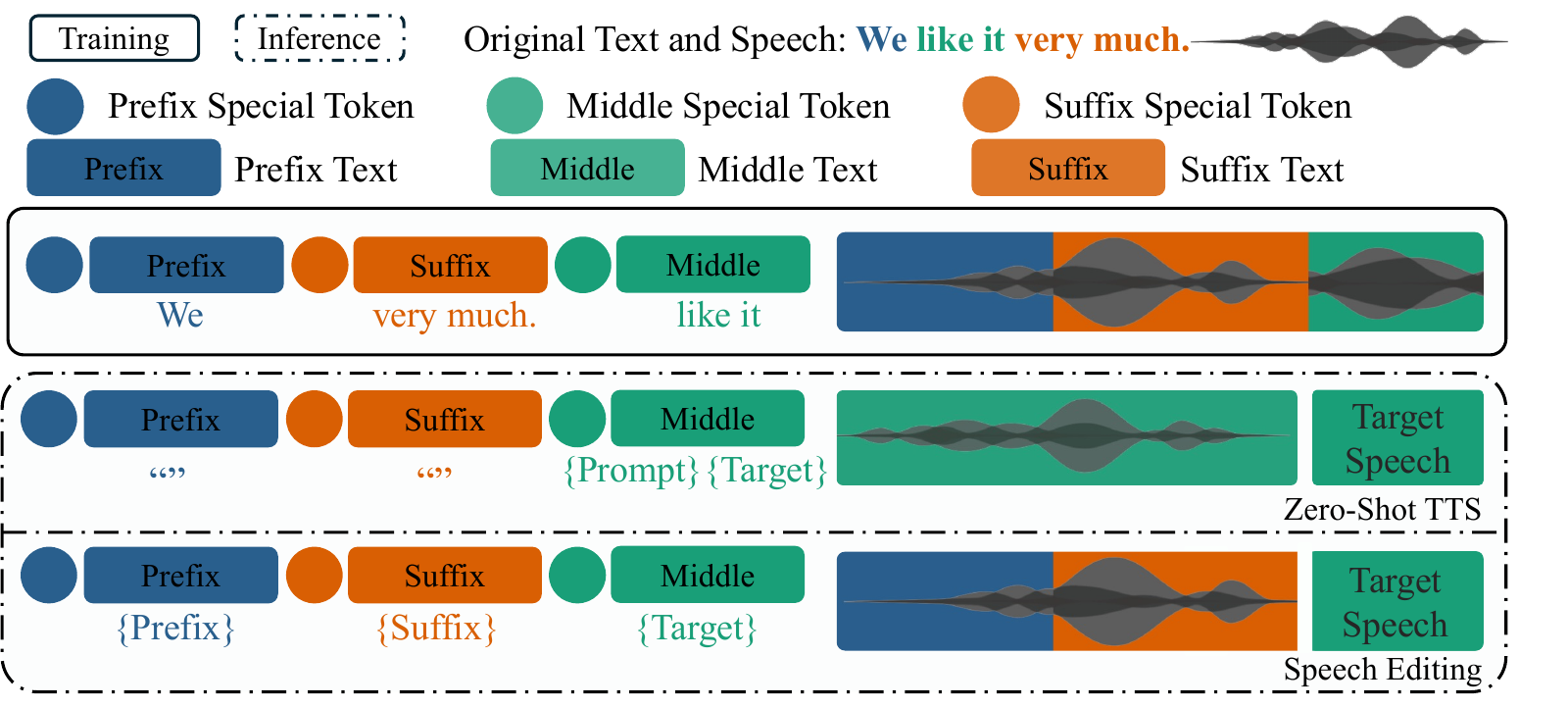}
    \caption{Illustration of Token Reordering}
    \label{fig:reorder}
    \vspace{-0.5cm}
\end{figure*}

VoiceCraft-X employs several token reordering steps, illustrated in Figure~\ref{fig:reorder}, to unify speech editing and synthesis. We assume that our training examples consist of utterance waveforms accompanied by time-aligned word transcriptions (we use the Montreal Forced Aligner (MFA)~\citep{mcauliffe2017montreal} in our work). During training, a text transcription is randomly segmented into prefix, middle, and suffix portions. These are then rearranged into a "prefix-suffix-middle" sequence, where the "middle" segment serves as the prediction target. Finally, the corresponding speech tokens for each segment are reordered identically based on the alignment timings. This ensures a monotonic alignment between the text and speech tokens, even when performing speech edits which require infilling tokens in the middle of the speech sequence. This rearrangement serves to mirror the use case in which a user wishes to modify some, but not all of the words in an utterance - by using this rearrangement, the model can be trained to predict the speech tokens within the middle of an utterance, conditioned on the preceding (prefix) and following (suffix) speech tokens in addition to the desired text transcription.

\subsection{Causal Masking and Delay Pattern}\label{sec:delay}
Following the token reordering, a learnable \textit{<MASK>} token is inserted at two locations within the text-speech input sequence: one \textit{<MASK>} token is inserted at the boundary between the prefix and suffix speech tokens, and a second \textit{<MASK>} token is placed between the suffix audio tokens and the middle (target) audio tokens. These tokens serve to inform the model of the boundaries between the segments.

During training, the model is tasked with autoregressively predicting all audio tokens: encompassing those in the prefix, suffix, and the middle (target) segments. This prediction is optimized using a standard language modeling objective, where the cross-entropy loss function is applied to every token in the sequence. By training the model to predict not only the target segment but also the known prefix and suffix segments, it receives gradients for every timestep, resulting in faster training.

%The audio tokens themselves are generated by a neural codec model, specifically the EnCodec~\citep{defossez2022high} model, which quantizes the audio waveform into multiple parallel streams or codebooks.
To model the $K$ parallel token sequences output by the EnCodec tokenizer autoregressively, we incorporate the ``Delay Pattern'' proposed by MusicGen~\citep{copet2023simple}. Instead of predicting all $K$ codebooks for a given audio timestep $t$ simultaneously or flattening all codebooks across all timesteps into one long sequence, delay patterning inserts a cumulative time delay of one timestep per RVQ layer to the EnCodec token sequences. As a result, the prediction for the speech token at codebook level $k$ at timestep $t$ can be conditioned on the model's predictions for codebook levels 1 through $k-1$ associated with the same timestep $t$.

%More formally, let an audio segment (e.g., prefix, suffix, or middle) be represented by a matrix of codec tokens $Y \in \mathbb{N}^{T \times K}$, where $T$ is the number of timesteps in the segment and $K$ is the number of codebooks. Applying the delay pattern rearranges this matrix into a single sequence $Z = (z_0, z_1, \dots, z_{T+K-2})$. Each element $z_j$ in this sequence is a vector of $K$ tokens, but these tokens are drawn from different original timesteps in $Y$. Specifically, for a given sequence step $j$ in the delayed sequence, the token for the $k$-th codebook is taken from timestep $j-k+1$ of the original audio segment $Y$. That is, $z_j = (Y_{j,1}, Y_{j-1,2}, \dots, Y_{j-K+1,K})$.

%To ensure that each $z_j$ contains $K$ valid tokens, especially at the beginning and end of the sequence where indices might go out of bounds, a special learnable \texttt{[empty]} token is used for padding. For instance, $Y_{t',k}$ is defined as \texttt{[empty]} if $t' < 0$ or $t' \ge T$. The same delay pattern is also applied to the \textit{<MASK>} tokens embedded in the audio spans.  This staggered arrangement helps the model capture dependencies both across codebooks and over time more effectively. See Figure~\ref{fig:architecture} for an illustration.

\subsection{Inference}
Figure~\ref{fig:reorder} shows how, at inference time, VoiceCraft-X performs speech editing and zero-shot text-to-speech by preparing an input sequence based on the "prefix-suffix-middle" reordering of text and speech tokens. The system then autoregressively generates the neural codec tokens for the target audio segment.%, leveraging the learned causal masking strategy and the delay pattern for multi-codebook generation.

\paragraph{Speech editing}
Let $T_P,\,A_P$ be the prefix text/audio, $T_S,\,A_S$ the suffix, and $T_M^{\text{new}}$ the user-supplied replacement text for the middle segment. The model input is the concatenation
\vspace{-0.25cm}
\begin{equation*}
\begin{split}
    T_P,\; T_S,\; T_M^{\text{new}},\
    \textit{<SPK>},\; A_P,\; \textit{<M>},\;  A_S,\; \textit{<M>},
\end{split}
\end{equation*}

where \textit{<SPK>} is a speaker embedding token and \textit{<M>} is the (learnable) mask token. The decoder predicts the middle-segment audio tokens $\hat A_M$, which we splice between $A_P$ and $A_S$ before decoding the entire sequence with the EnCodec decoder network to create a seamless edit.

\paragraph{Zero-shot TTS}
If a prompt text ($T_{prompt}$) and its corresponding prompt speech are provided, we concatenate the prompt text and the target text ($T_{target}$) to form the middle text segment, and a speaker embedding is extracted from the prompt speech. If no such prompt is provided, we set the prompt text ($T_{prompt}$) to empty and randomly generate a speaker embedding. %\david{What is the use case for the model with an empty prompt? Do we ever do any experiments with this configuration?} 
The final input is as follows:

\vspace{-0.45cm}
\begin{equation*}
\begin{split}
    & T_P,\; T_S,\; T_{prompt},\; T_{target},\; \\
    & \textit{<SPK>},\; A_P,\; \textit{<M>},\; A_S,\; \textit{<M>},\; A_{prompt},
\end{split}
\end{equation*}
where $T_P=T_S=\varnothing$, $A_P=A_S=\varnothing$, and $T_{prompt}=A_{prompt}=\varnothing$ if no prompt is provided.

\section{Experiments}
\subsection{Setup}\label{sec:experiments_setup}
\paragraph{Training Dataset.} We combined speech data across public datasets over 11 languages, amounting to a total of approximately 32K hours (detailed statistics provided in Appendix~\S\ref{app:dataset}). The sampling rate for all audio is 16 kHz. Audio segments longer than 25 seconds were discarded. For MLS dataset~\citep{pratap2020mls}, misalignment issues were particularly prominent, with approximately 20\% of samples having extra or missing words in the transcript at the beginning or end. We found that this negatively impacted model performance for English, and subsequently removed utterances whose transcriptions differed significantly from those produced by the Whisper~\citep{radford2023robust} model. While we found similar problems with the non-English European language data in MLS, we anecdotally observed better performance on those languages without performing this filtering. We speculate that this is due to the fact that the amount of available training data for those languages is already relatively low, and the performance improvements brought by the additional training data outweigh the detriments brought by transcription noise.

\paragraph{Evaluation Dataset.} For evaluating Text-to-Speech (TTS) performance, we curated an evaluation dataset from several established benchmarks. For English, we utilized the Seed-TTS test-en set~\citep{anastassiou2024seed} (1088 samples sourced from Common Voice~\citep{ardila2019common}). For Mandarin, we employed the Seed-TTS test-zh set (2020 samples from DiDiSpeech~\citep{guo2021didispeech}). Korean and Japanese evaluations were conducted using 200 randomly selected samples from KsponSpeech~\citep{bang2020ksponspeech} and KokoroSpeech~\citep{Iida2021Kokoro}, respectively. For the remaining seven languages supported by our model (Spanish, French, German, Dutch, Italian, Portuguese, and Polish), we randomly selected 100 samples for each language from their corresponding Multilingual LibriSpeech (MLS)~\citep{pratap2020mls} test sets. To evaluate speech editing, we randomly selected 100-300 samples per language from these TTS test datasets and then utilized Gemini~\citep{team2023gemini} to perform insertion, deletion, or substitution operations on the textual portions of these samples, with specific details available in the appendix~\S\ref{app:editing_dataset}. We conducted subjective evaluation over a subset of languages (English, Chinese, French, Italian, Portuguese, and Spanish) using a random subset of the evaluation set: 40 English samples, 50 Chinese, and 20 for others.

\paragraph{Training.}
Our model utilizes Encodec~\citep{defossez2022high} as the speech tokenizer. We retrain the model with some modifications, namely using 4 Residual Vector Quantization (RVQ) codebooks, each containing 2048 entries, and a framerate of 50Hz on audio recorded at 16 kHz. We retrain the model with our multilingual speech data. Other than those, the training process adheres to the methodology outlined in the work by~\citep{defossez2022high}. Additional configuration specifics can be found in Section~\S\ref{app:encodec}. To combine the parallel speech tokens when using them as input to the Transformer LM, at each timestep we sum the embeddings of the tokens across the four codebooks.

We use Qwen3-0.6B-Base as both the text tokenizer and the Transformer LM backbone (details are provided in Appendix~\ref{app:qwen3_details}). The outputs from the final Transformer layer are then projected into four distinct linear layers, each producing the logits for one of the codec tokens. The model comprises 613 million total parameters (457 million excluding embeddings). The codebook weights $\boldsymbol{\alpha}$ are set to $(1.0, 0.8, 0.6, 0.4)$, influencing the contribution of each codebook during training (as further detailed in our loss formulation~\S\ref{app:loss_design}). For model training, we employ the AdamW optimizer~\citep{loshchilov2017decoupled} with a learning rate of $4 \times 10^{-3}$, $\beta_1=0.9$, $\beta_2=0.999$, an epsilon of $1 \times 10^{-6}$, and a weight decay of $0.01$. A learning rate scheduler is utilized, featuring a linear warm-up for the initial $50K$ steps, followed by a linear decay for the remainder of the $5,000K$ total training steps. Gradient accumulation is performed over $8$ micro-batches. The training of the multilingual VoiceCraft-X model took approximately one week on 16 NVIDIA A100 40GB GPUs.

\paragraph{Inference}
Figure~\ref{fig:reorder} shows how, at inference time, VoiceCraft-X performs speech editing and zero-shot text-to-speech by preparing an input sequence based on the "prefix-suffix-middle" reordering of text and speech tokens; the model then autoregressively predicts the corresponding neural codec tokens for the target audio segment. Notably, the token reordering mechanism significantly enhances inference stability. This largely prevents repeating token loops, an issue in the original VoiceCraft which could cause artifacts (e.g., excessive silences) and required multi-sample filtering. Consequently, VoiceCraft-X reliably generates high-quality speech in a single pass without needing this filtering step. In all experiments, we employ nucleus sampling~\citep{holtzman2019curious} with $TopK=20, TopP=1.0$, and a temperature of 1.

\paragraph{Baselines.}
For the English and Chinese Zero-shot TTS tasks, we compared our model with FireRedTTS~\citep{guo2024fireredtts}, MaskGCT~\citep{wang2024maskgct}, F5-TTS~\citep{chen2024f5}, CosyVoice~\citep{du2024cosyvoice1}, and CosyVoice 2~\citep{du2024cosyvoice2}. For English, we also included VoiceCraft~\citep{peng2024voicecraft} in our comparison. For the remaining languages, we benchmarked our model against the multilingual XTTS~\citep{casanova2024xtts} model, considering both its v1 and v2 versions. For speech editing, we compared VoiceCraft-X with the original VoiceCraft~\citep{peng2024voicecraft} model on English.
% considering the training data scale of our model; VoiceCraft utilizes a comparable amount of data, unlike many other contemporary editing systems that are trained on significantly larger datasets.

\begin{table*}[t]
\vspace{-0.25cm}
\centering
\small
\caption{Zero-Shot TTS performance across different models and languages. \textsuperscript{\textdaggerdbl}\textit{Training Hours} for XTTS-v2 may be an underestimation as the model is continuously updated and specific training data has not been fully disclosed. "-" indicates data not available or not applicable. *For Chinese, Korean and Japanese, figures in the WER columns represent Character Error Rate (CER). \textsuperscript{\textdagger}Scores reported in baseline papers. }
\label{tab:tts_comparison}
\vspace{-0.3cm}
\begingroup 
\setlength{\tabcolsep}{3.5pt}
\begin{tabular}{@{}c ccccc ccccc@{}}
\toprule
& \multicolumn{5}{c}{Chinese*} & \multicolumn{5}{c}{English} \\
\cmidrule(lr){2-6} \cmidrule(lr){7-11}
& Train (hrs) & WER & SIM-o & CMOS & SMOS & Train (hrs) & WER & SIM-o & CMOS & SMOS \\
\midrule
Ground Truth & \text{-} & 1.25 & 0.75 & 0.0 & 3.38 & \text{-} & 2.14 & 0.73 & 0.0 & 3.36\\
\midrule
MaskGCT~\citep{wang2024maskgct} & 49.9K & 2.27\textsuperscript{\textdagger} & \textbf{0.77}\textsuperscript{\textdagger} & \text{-} & \text{-} & 46.8K & 2.62\textsuperscript{\textdagger} & \textbf{0.72}\textsuperscript{\textdagger} & \text{-} & \text{-} \\
F5-TTS~\citep{chen2024f5} & 49.9K & 1.56\textsuperscript{\textdagger} & 0.76\textsuperscript{\textdagger} & \text{-} & \text{-} & 46.8K & \textbf{1.83}\textsuperscript{\textdagger} & 0.67\textsuperscript{\textdagger} & \text{-} & \text{-} \\
FireRedTTS~\citep{guo2024fireredtts} & 110K & \textbf{1.21} & 0.65 & -0.28 & 2.82 & 40K & 9.08 & 0.45 & 0.27 & 2.97 \\
CosyVoice~\citep{du2024cosyvoice1} & 130K & 3.49 & 0.75 & \textbf{0.18} & 3.64 & 30K & 3.89 & 0.64 & 0.50 & 3.48 \\
CosyVoice 2~\citep{du2024cosyvoice2} & 130K & 1.35 & 0.75 & -0.01 & \textbf{3.86} & 30K & 2.69 & 0.65 & 0.59 & \textbf{3.69} \\
\midrule
VoiceCraft~\citep{peng2024voicecraft} & \text{-} & \text{-} & \text{-} & \text{-} & \text{-} & 9K & 5.28 & 0.51 & 0.44 & 3.27 \\
VoiceCraft-X & 5K & 3.29 & 0.68 & -0.39 & 2.94 & 14.5K & 4.20 & 0.54& \textbf{0.63} & 3.43 \\
\bottomrule
\end{tabular}
\endgroup

\begin{tabular}{@{}c ccc ccc ccc@{}}
\toprule
& \multicolumn{3}{c}{Korean*} & \multicolumn{3}{c}{Japanese*} & \multicolumn{3}{c}{Dutch} \\
\cmidrule(lr){2-4} \cmidrule(lr){5-7} \cmidrule(lr){8-10}
& Train (hrs)  & {WER} & {SIM-o} & Train (hrs)  & {WER} & {SIM-o} & Train (hrs)  & {WER} & {SIM-o} \\
\midrule
Ground Truth & \text{-} & 8.89 & \text{-} & \text{-} & 9.72 & 0.79 & \text{-} & 9.54 & 0.65 \\
\midrule
XTTS-v1& \text{-} & \text{-} & \text{-} & \text{-} & \text{-} & \text{-} & \text{-} & 78.17 & 0.41 \\
XTTS-v2 & 539\textsuperscript{\textdaggerdbl} & 40.89 & \textbf{0.62} & 57\textsuperscript{\textdaggerdbl} & \textbf{11.61} & 0.64 & 74\textsuperscript{\textdaggerdbl} & \textbf{12.62} & 0.59 \\
\midrule
VoiceCraft-X & 832 & \textbf{31.11} & 0.56 & 3489 & 15.09 & \textbf{0.66} & 2147 & 16.28 & \textbf{0.61} \\
\bottomrule
\end{tabular}

\begin{tabular}{@{}c ccc ccc ccc@{}}
\toprule
& \multicolumn{3}{c}{Italian} & \multicolumn{3}{c}{Portuguese} & \multicolumn{3}{c}{Polish} \\
\cmidrule(lr){2-4} \cmidrule(lr){5-7} \cmidrule(lr){8-10}
 & Train (hrs)  & {WER} & {SIM-o} & Train (hrs)  & {WER} & {SIM-o} & Train (hrs)  & {WER} & {SIM-o} \\
\midrule
Ground Truth & \text{-} & 9.48 & 0.68 & \text{-} & 8.75 & 0.69 & \text{-} & 8.81 & 0.72 \\
\midrule
XTTS-v1 & \text{-} & 73.12 & 0.32 & \text{-} & 48.93 & 0.33 & \text{-} & 96.15 & 0.41 \\
XTTS-v2 & 1297\textsuperscript{\textdaggerdbl} & 15.52 & \textbf{0.56} & 2387\textsuperscript{\textdaggerdbl} & \textbf{13.48} & \textbf{0.58} & 199\textsuperscript{\textdaggerdbl} & \textbf{9.47} & \textbf{0.62} \\
\midrule
VoiceCraft-X & 294 & \textbf{15.46} & 0.54 & 223 & 22.57 & 0.56 & 139 & 24.80 & 0.61 \\
\bottomrule
\end{tabular}

\begin{tabular}{@{}c ccc ccc ccc@{}}
\toprule
& \multicolumn{3}{c}{French} & \multicolumn{3}{c}{German} & \multicolumn{3}{c}{Spanish} \\
\cmidrule(lr){2-4} \cmidrule(lr){5-7} \cmidrule(lr){8-10}
 & Train (hrs)  & {WER} & {SIM-o} & Train (hrs)  & {WER} & {SIM-o} & Train (hrs)  & {WER} & {SIM-o} \\
\midrule
Ground Truth & \text{-} & 6.09 & 0.68 & \text{-} & 6.64 & 0.69 & \text{-} & 4.87 & 0.73 \\
\midrule
XTTS-v1 & \text{-} & 38.34 & 0.35 & \text{-} & 11.37 & 0.35 & \text{-} & 20.84 & 0.37 \\
XTTS-v2 & 2216\textsuperscript{\textdaggerdbl} & \textbf{5.45} & 0.58 & 3584\textsuperscript{\textdaggerdbl} & 16.50 & 0.59 & 1514\textsuperscript{\textdaggerdbl} & 8.11 & 0.58 \\
\midrule
VoiceCraft-X & 1338 & 13.22 & \textbf{0.59} & 3405 & \textbf{8.19} & \textbf{0.60} & 1191 & \textbf{4.67} & \textbf{0.63} \\
\bottomrule
\end{tabular}
\vspace{-0.4cm}
\end{table*}

\paragraph{Metrics.}
We used a combination of subjective and objective measures. Objectively, we use Word Error Rate (WER) as an automatic proxy for the intelligibility of the synthesized speech; this is calculated using Paraformer-zh~\citep{gao2023funasr} for Chinese and Whisper-large-v3~\citep{radford2023robust} for other languages. Additionally, speaker similarity (SIM-o) is objectively measured by computing the cosine similarity of speaker embeddings, which are extracted from both the generated and original target speech using a WavLM-based speaker verification model~\citep{chen2022wavlm}. Subjective evaluations involved human annotators (see Appendix~\ref{app:subjective} for details) who provide Comparative Mean Opinion Scores (CMOS) and Similarity Mean Opinion Scores (SMOS) for TTS, and Naturalness Mean Opinion Scores (NMOS) and Intelligibility Mean Opinion Scores (IMOS) for speech editing. For CMOS, evaluators assess the naturalness of the synthesized speech in comparison to the ground truth, while for SMOS, they directly score the similarity between the synthesized speech and the initial speech prompt. For NMOS and IMOS, evaluators respectively assess the naturalness and intelligibility of the synthesized and original speech.

\subsection{Zero-Shot TTS}
We evaluated VoiceCraft-X's zero-shot TTS performance across 11 languages, and the results are shown in Table~\ref{tab:tts_comparison}. For Chinese, VoiceCraft-X was trained on a modest $5K$ hours of data, a fraction of that used by leading models (often exceeding $50K$ hours). Consequently, while its CER of 3.29 was higher than these specialized models, this was achieved with substantially less data, and its speaker similarity and subjective scores reflected this data disparity. In English, VoiceCraft-X, trained on $14K$ hours, showed marked improvements over its predecessor, VoiceCraft, reducing its WER from 5.28 to 4.37 and enhancing SIM-o from 0.51 to 0.54. Critically, its CMOS score of 0.63\footnote{The generally higher English CMOS scores likely resulted from using Seed-TTS test set as prompts with atypical, exaggerated intonation (not standard read speech).} was the highest among compared models, indicating superior perceived naturalness. While some models trained on significantly larger datasets achieved lower WERs, VoiceCraft-X's subjective quality in English was highly competitive.

For the remaining nine languages, VoiceCraft-X, compared to XTTS (versions v1 and v2), showed strong overall performance with varying focuses. VoiceCraft-X particularly excelled in European languages like German (WER significantly better than XTTS-v2 by over 50\%), Spanish (WER over 40\% better than XTTS-v2 and below the ground truth), and Italian (higher data efficiency), as well as in Korean (CER reduced by over 20\%). However, in languages such as Japanese and Dutch, or for those where VoiceCraft-X had considerably less training data like Portuguese and Polish, XTTS-v2 achieved lower error rates. Nevertheless, VoiceCraft-X was often favored by evaluators for its better speaker similarity, naturalness, and intelligibility. (Further results are in the appendix~\S\ref{app:subjective}).

\subsection{Transfer Learning for Multilingual TTS}\label{sec:insights_transfer}
\begin{table*}[htp]
\vspace{-0.3cm}
\centering
\caption{Cross-lingual transfer learning performance on zero-shot TTS task. Comparison of fine-tuning from different pre-trained models versus training from scratch for various target languages. Character Error Rate (CER) for Korean and Japanese, indicated by *. "-" indicates data not available or not applicable.}
\label{tab:cross_lingual_transfer}
\small
\setlength{\tabcolsep}{4pt} % Reduce inter-column spacing
\begin{tabular}{@{}c c cc cc cc cc cc@{}}
\toprule
\multirow{2}{*}{\textbf{Language}} & \multirow{2}{*}{\textbf{\#Hours}} & \multicolumn{2}{c}{\textbf{Multilingual}} & \multicolumn{2}{c}{\textbf{from Scratch}} & \multicolumn{2}{c}{\textbf{from English}} & \multicolumn{2}{c}{\textbf{from Chinese/Japanese}} & \multicolumn{2}{c}{\textbf{from Multilingual}} \\
\cmidrule(lr){3-4} \cmidrule(lr){5-6} \cmidrule(lr){7-8} \cmidrule(lr){9-10} \cmidrule(lr){11-12}
& & WER & SIM-o & WER & SIM-o & WER & SIM-o & WER & SIM-o & WER & SIM-o \\
\midrule
Korean* & 832 & 31.11 & \textbf{0.56}   & 45.79 & 0.51 & 42.10 & 0.54 & 49.11/42.08 & 0.50/0.52 & 41.36 & 0.53 \\
Japanese* & 3489 & \textbf{15.09} & 0.66 & 22.36  & 0.62 & -      & -     & 36.18         & 0.61         & 19.35 & \textbf{0.67} \\
Spanish & 1191 & 4.67	& \textbf{0.63} & 7.08  & 0.38 & 4.54  & 0.47  & -             & -             & \textbf{3.30}  & 0.52 \\
French &  1338 & 13.22	& \textbf{0.60} & 18.85 & 0.43 & \textbf{12.50} & 0.49 & -             & -             & 16.39 & 0.53 \\
German & 3405 & 8.19	& \textbf{0.60} & 6.43   & 0.43 & \textbf{5.93}   & 0.50 & -             & -             & 7.25   & 0.53 \\
Dutch & 2147 & 16.28 & \textbf{0.61} & 16.85 & 0.37 & 16.02 & 0.35 & -             & -             & \textbf{11.78} & 0.46  \\
Italian & 294 & 15.46 & \textbf{0.54} & 142.30& 0.22 & 13.97 & 0.36 & -             & -             & \textbf{13.93} & 0.46 \\
Portuguese  & 223 & 22.57 & \textbf{0.56} & 91.89 & 0.26 & 15.87 & 0.46 & -             & -             & \textbf{14.74} & 0.55 \\
Polish & 139 & 24.80 & \textbf{0.61} & 163.08& 0.25 & 20.73 & 0.46 & -             & -             & \textbf{19.47} & 0.55 \\
\bottomrule
\end{tabular}
\vspace{-0.4cm}
\end{table*}

To explore the benefits of multilingual training, especially for lower-resource languages, we fine-tuned \textit{monolingual} models on individual languages starting from different pre-trained checkpoints, comparing these against training from scratch and the multilingual model (detailed in Table~\ref{tab:cross_lingual_transfer}). 

The universal advantage of pre-training over ``from Scratch'' models is paramount, especially for languages with limited data. For instance, Italian (294 hours) and Polish (139 hours) saw their WERs plummet from over 140 and 160 to under 14 and 20 respectively, demonstrating pre-training's crucial role in transferring foundational knowledge and overcoming data scarcity. Even higher-resource languages like Spanish, French and German benefited significantly. Fine-tuning from an English model initialization proved highly effective for European languages (Germanic, Romance, Slavic), leveraging linguistic similarities and robust acoustic modeling, with gains particularly vital for low-data scenarios (Italian, Portuguese, Polish). Korean showed better CER with a Japanese checkpoint (42.08) than Chinese (49.11), aligning with typological closeness. Conversely, Japanese experienced negative transfer from Chinese (CER 36.18 vs. 22.36 from scratch).

Furthermore, fine-tuning from the ``multilingual checkpoint'' frequently yielded superior WER/CER compared to an English-only checkpoint for a range of languages including Spanish, Dutch, Italian, Portuguese, Polish, and Japanese. This advantage held across varying data volumes (e.g., Polish 139 hours, Japanese 3489 hours), suggesting that pre-training on a diverse linguistic set fosters more generalized and transferable representations than exposure to English alone, capturing a broader array of phonetic and prosodic patterns.

Finally, the original multilingual model's speaker similarity is significantly higher than models fine-tuned from other checkpoints for nearly all languages. This indicates that joint training on diverse linguistic data, leveraging collective data volume, allows the model to disentangle speaker-specific characteristics from language-specific features. This robust performance across varied languages suggests it learns a more abstract, shared representation space for speech, facilitating both high-fidelity synthesis and strong cross-lingual capabilities. While fine-tuning on single language data may impact this disentanglement ability, as evidenced by SIM-o drops in many such cases.

\subsection{Speech Editing}
\begin{table}[htp]
  \centering
  \caption{Performance on English speech editing.}
  \label{tab:speech-editing-en}
  {
    \renewcommand{\arraystretch}{0.85}
    \begin{tabular}{c@{\hspace{1em}}c@{\hspace{1em}}c@{\hspace{1em}}c}
      \toprule
       & WER & NMOS & IMOS \\
      \midrule
      Original & 2.42 & 3.78 & 3.79 \\
      \midrule
      VoiceCraft & 5.99 & \textbf{3.87} & \textbf{3.87} \\
      VoiceCraft-X & \textbf{5.62} & 3.68 & 3.79 \\
      \bottomrule
    \end{tabular}
  }
\vspace{-0.25cm}
\end{table}

For English speech editing (Table~\ref{tab:speech-editing-en}), VoiceCraft-X demonstrated a better Word Error Rate (WER) than VoiceCraft. Both models produced edited speech that listeners found to be highly natural (NMOS) and intelligible (IMOS), comparable to the original recordings. VoiceCraft's slightly higher scores in these subjective tests are not surprising, given its monolingual English focus, especially considering both models have similar parameter counts and amounts of English training data.

\begin{table}[htbp]
  \centering
  \footnotesize
  \caption{Subjective performance on speech editing.}
  \label{tab:mos_speech_editing}
  \begin{tabular}{@{}ccccc@{}}
    \toprule
    & \multicolumn{2}{c}{Original} & \multicolumn{2}{c}{Edited} \\
    \cmidrule(r){2-3} \cmidrule(l){4-5} 
    & NMOS & IMOS & NMOS & IMOS \\
    \midrule
    \textbf{French} & 3.62 & 4.10 & 3.13 & 3.60 \\
    \textbf{Italian} & 4.38 & 4.78 & 3.77 & 4.28 \\
    \textbf{Portuguese} & 4.42 & 4.98 & 2.63 & 3.78 \\
    \textbf{Spanish} & 3.80 & 3.93 & 3.58 & 3.78 \\
    \bottomrule
  \end{tabular}
\vspace{-0.5cm}
\end{table}

For multilingual speech editing in other languages—a capability where comparative baselines are notably scarce as most models do not support multilingual editing—we conducted subjective MOS evaluations. These evaluations focused on a subset of languages (French, Italian, Portuguese, and Spanish) for which MTurk annotators were available, with results presented in Table~\ref{tab:mos_speech_editing}. The evaluations demonstrate VoiceCraft-X's effective performance in this challenging scenario. While naturalness (NMOS) scores for edited speech are, as anticipated, lower than the original recordings, intelligibility (IMOS) remains high across these languages. Particularly for Spanish and Italian, where edited NMOS and IMOS scores closely matched the original audio, these findings underscore VoiceCraft-X's significant and unique capability for coherent, comprehensible multilingual speech editing.

%demonstrated its capability in speech editing as detailed in Table~\ref{tab:speech-editing}. The WER for edited speech was consistently higher than the WER of the original audio, which is expected as speech synthesis, especially in an editing context, can introduce some level of error. Despite these increases, the model successfully performed editing operations across all listed languages, showcasing its multilingual editing functionality. Additional subjective evaluation for a subset of these languages can be found in the appendix~\S\ref{app:subjective}.

% \begin{center}
%   \captionof{table}{WER on multilingual speech editing.}
%   \label{tab:speech-editing}
%   \small
%   \setlength{\tabcolsep}{4pt}
%   \begin{tabular}{ccccccccccc}
%     \toprule
%      & {ZH*} & {KO*} & {JA*} & {NL} & {IT} \\
%     \midrule
%     Original & 1.48 & 18.26 & 18.24 & 7.28 & 8.82 \\
%     Edited   & 4.31 & 39.73 & 19.58 & 18.61 & 18.10 \\
%     \midrule
%     & {PT} & {PL} & {FR} & {DE} & {ES} \\\midrule
%     Original & 7.81 & 7.92 & 3.94 & 6.15 & 4.29 \\
%     Edited & 21.14 & 14.79 & 11.80 & 11.08 & 10.39 \\
%     \bottomrule
%   \end{tabular}
% \end{center}

\section{Conclusion}
We present VoiceCraft-X, an autoregressive neural codec language model that successfully unifies multilingual speech editing and Text-to-Speech (TTS) synthesis. Leveraging the Qwen3 LLM and a novel token reordering strategy, VoiceCraft-X supports eleven languages, producing high-quality, natural-sounding speech. Our model demonstrates robust performance across diverse conditions and shows that a unified framework can effectively advance both speech editing and synthesis in multilingual contexts, even with limited data for some languages. This work underscores the potential of autoregressive models for complex, real-world speech generation tasks.

\clearpage
\section*{Limitations}
One key limitation is the scale of our training data. Although VoiceCraft-X performs well with approximately 32,578 hours across eleven languages, this is notably less than some state-of-the-art models. This comparative data scarcity, particularly for lower-resource languages in our set, may limit the model's capacity to capture the full spectrum of speech nuances as effectively as systems trained on more extensive datasets.

Secondly, while the model's multilingual support is a core feature, its current reach of eleven languages (with around 20-30 explored internally) only scratches the surface of global linguistic diversity. Expanding coverage to more languages, especially under-resourced ones, remains a significant challenge that would require substantial data curation and potential model adaptations to address varied linguistic features.

Finally, further investigation into model size scalability is also warranted. The current VoiceCraft-X utilizes the Qwen3-0.6B architecture; exploring larger model variants could unlock enhanced learning capabilities and higher fidelity in speech synthesis and editing. Systematically assessing different model sizes is crucial for optimizing the balance between performance improvements and computational demands.

\section*{Ethical Implications}
The development of advanced speech models like VoiceCraft-X, which possesses strong zero-shot voice cloning and multilingual editing capabilities, carries significant ethical responsibilities. We acknowledge the potential for misuse of this technology. Malicious actors could exploit it for unauthorized voice cloning, impersonation, the creation of convincing deepfakes for fraudulent purposes, or the generation of misinformation and propaganda. These risks are particularly pronounced given the model's ability to operate across eleven languages, broadening the potential scope for misuse on a global scale.

The zero-shot nature of VoiceCraft-X lowers the barrier to entry for creating high-fidelity synthetic audio, making it accessible to a wider range of actors beyond those with specialized technical expertise. This accessibility amplifies the dual-use nature of the technology; while it empowers creativity and accessibility, it also provides a powerful tool for deception.

We recognize that technical solutions alone are insufficient to address these societal challenges. The proliferation of convincing synthetic media necessitates a broader, collaborative effort involving researchers, platform companies, policymakers, and the public to develop new norms, regulations, and educational initiatives around the responsible creation and consumption of digital content.

To mitigate these risks, we are committed to a responsible release of our model and code. We strongly advocate for the research community to explore and develop robust safeguards, such as audio watermarking and detection tools, to help distinguish between authentic and synthesized audio. Such advancements are crucial for building a safer information ecosystem, \textit{but are only possible if open-source versions of these models are available for researchers to utilize.} Our release will be accompanied by strict intended-use guidelines and a license that explicitly prohibits malicious applications, such as impersonating public figures or private individuals without their explicit consent. We believe that by fostering an open yet cautious approach, we can encourage further research into safety measures while providing a valuable tool for beneficial applications and advancing the field of speech technology responsibly.

\section*{Acknowledgments}
This work was supported by Amazon.com, PO No. 2D-16003984 through the Amazon-UT Austin HUB. We thank Sanyuan Chen, Zhikang Niu, Chen Yang, Tianrui Wang, Yushen Chen, Yifan Yang, Xie Chen for their constructive feedback.

\bibliography{custom}

@article{borsos2022speechpainter,
  title={Speechpainter: Text-conditioned speech inpainting},
  author={Borsos, Zal{\'a}n and Sharifi, Matt and Tagliasacchi, Marco},
  journal={arXiv preprint arXiv:2202.07273},
  year={2022}
}

@inproceedings{le2023voicebox,
  title={Voicebox: Text-guided multilingual universal speech generation at scale},
  author={Le, Matthew and Vyas, Apoorv and Shi, Bowen and Karrer, Brian and Sari, Leda and Moritz, Rashel and Williamson, Mary and Manohar, Vimal and Adi, Yossi and Mahadeokar, Jay and others},
  booktitle={Proc. NeurIPS},
  year={2023}
}

@article{peng2024voicecraft,
  title={Voicecraft: Zero-shot speech editing and text-to-speech in the wild},
  author={Peng, Puyuan and Huang, Po-Yao and Li, Shang-Wen and Mohamed, Abdelrahman and Harwath, David},
  journal={arXiv preprint arXiv:2403.16973},
  year={2024}
}

@article{chen2024f5,
  title={F5-tts: A fairytaler that fakes fluent and faithful speech with flow matching},
  author={Chen, Yushen and Niu, Zhikang and Ma, Ziyang and Deng, Keqi and Wang, Chunhui and Zhao, Jian and Yu, Kai and Chen, Xie},
  journal={arXiv preprint arXiv:2410.06885},
  year={2024}
}

@article{wang2024maskgct,
  title={Maskgct: Zero-shot text-to-speech with masked generative codec transformer},
  author={Wang, Yuancheng and Zhan, Haoyue and Liu, Liwei and Zeng, Ruihong and Guo, Haotian and Zheng, Jiachen and Zhang, Qiang and Zhang, Xueyao and Zhang, Shunsi and Wu, Zhizheng},
  journal={arXiv preprint arXiv:2409.00750},
  year={2024}
}

@misc{qwen3,
    title  = {Qwen3},
    url    = {https://qwenlm.github.io/blog/qwen3/},
    author = {Qwen-Team},
    month  = {April},
    year   = {2025}
}

@article{koizumi2023libritts,
  title={Libritts-r: A restored multi-speaker text-to-speech corpus},
  author={Koizumi, Yuma and Zen, Heiga and Karita, Shigeki and Ding, Yifan and Yatabe, Kohei and Morioka, Nobuyuki and Bacchiani, Michiel and Zhang, Yu and Han, Wei and Bapna, Ankur},
  journal={arXiv preprint arXiv:2305.18802},
  year={2023}
}

@article{chen2021gigaspeech,
  title={Gigaspeech: An evolving, multi-domain asr corpus with 10,000 hours of transcribed audio},
  author={Chen, Guoguo and Chai, Shuzhou and Wang, Guanbo and Du, Jiayu and Zhang, Wei-Qiang and Weng, Chao and Su, Dan and Povey, Daniel and Trmal, Jan and Zhang, Junbo and others},
  journal={arXiv preprint arXiv:2106.06909},
  year={2021}
}

@article{pratap2020mls,
  title={Mls: A large-scale multilingual dataset for speech research},
  author={Pratap, Vineel and Xu, Qiantong and Sriram, Anuroop and Synnaeve, Gabriel and Collobert, Ronan},
  journal={arXiv preprint arXiv:2012.03411},
  year={2020}
}

@article{ma2024wenetspeech4tts,
  title={Wenetspeech4tts: A 12,800-hour mandarin tts corpus for large speech generation model benchmark},
  author={Ma, Linhan and Guo, Dake and Song, Kun and Jiang, Yuepeng and Wang, Shuai and Xue, Liumeng and Xu, Weiming and Zhao, Huan and Zhang, Binbin and Xie, Lei},
  journal={arXiv preprint arXiv:2406.05763},
  year={2024}
}

@article{du2018aishell,
  title={Aishell-2: Transforming mandarin asr research into industrial scale},
  author={Du, Jiayu and Na, Xingyu and Liu, Xuechen and Bu, Hui},
  journal={arXiv preprint arXiv:1808.10583},
  year={2018}
}

@article{bang2020ksponspeech,
  title={Ksponspeech: Korean spontaneous speech corpus for automatic speech recognition},
  author={Bang, Jeong-Uk and Yun, Seung and Kim, Seung-Hi and Choi, Mu-Yeol and Lee, Min-Kyu and Kim, Yeo-Jeong and Kim, Dong-Hyun and Park, Jun and Lee, Young-Jik and Kim, Sang-Hun},
  journal={Applied Sciences},
  year={2020},
}

@article{yin2023reazonspeech,
  title={ReazonSpeech: A free and massive corpus for Japanese ASR},
  author={Yin, Yue},
  year={2023}
}

@inproceedings{oliveira2023cml,
  title={Cml-tts: A multilingual dataset for speech synthesis in low-resource languages},
  author={Oliveira, Frederico S and Casanova, Edresson and Junior, Arnaldo Candido and Soares, Anderson S and Galv{\~a}o Filho, Arlindo R},
  booktitle={Proc. TSD},
  year={2023}
}

@misc{magicdata,
  author       = {{Magic Data}},
  title        = {MAGICDATA Mandarin Chinese Read Speech Corpus},
  year         = {2019}
}

@inproceedings{copet2023simple,
  title={Simple and controllable music generation},
  author={Copet, Jade and Kreuk, Felix and Gat, Itai and Remez, Tal and Kant, David and Synnaeve, Gabriel and Adi, Yossi and D{\'e}fossez, Alexandre},
  booktitle={Proc. NeurIPS},
  year={2023}
}

@article{defossez2022high,
  title={High fidelity neural audio compression},
  author={D{\'e}fossez, Alexandre and Copet, Jade and Synnaeve, Gabriel and Adi, Yossi},
  journal={arXiv preprint arXiv:2210.13438},
  year={2022}
}

@inproceedings{saeki2024extending,
  title={Extending multilingual speech synthesis to 100+ languages without transcribed data},
  author={Saeki, Takaaki and Wang, Gary and Morioka, Nobuyuki and Elias, Isaac and Kastner, Kyle and Rosenberg, Andrew and Ramabhadran, Bhuvana and Zen, Heiga and Beaufays, Fran{\c{c}}oise and Shemtov, Hadar},
  booktitle={Proc. ICASSP},
  year={2024},
}

@inproceedings{liu2025voxpopulitts,
  title={VoxpopuliTTS: a large-scale multilingual TTS corpus for zero-shot speech generation},
  author={Liu, Wenrui and Bai, Jionghao and Cheng, Xize and Zuo, Jialong and Jiang, Ziyue and Ji, Shengpeng and Fang, Minghui and Yang, Xiaoda and Yang, Qian and Zhao, Zhou},
  booktitle={Proc. COLING},
  year={2025}
}

@article{liao2024fish,
  title={Fish-speech: Leveraging large language models for advanced multilingual text-to-speech synthesis},
  author={Liao, Shijia and Wang, Yuxuan and Li, Tianyu and Cheng, Yifan and Zhang, Ruoyi and Zhou, Rongzhi and Xing, Yijin},
  journal={arXiv preprint arXiv:2411.01156},
  year={2024}
}

@inproceedings{radford2023robust,
  title={Robust speech recognition via large-scale weak supervision},
  author={Radford, Alec and Kim, Jong Wook and Xu, Tao and Brockman, Greg and McLeavey, Christine and Sutskever, Ilya},
  booktitle={Proc. ICML},
  year={2023},
}

@article{casanova2024xtts,
  title={Xtts: a massively multilingual zero-shot text-to-speech model},
  author={Casanova, Edresson and Davis, Kelly and G{\"o}lge, Eren and G{\"o}knar, G{\"o}rkem and Gulea, Iulian and Hart, Logan and Aljafari, Aya and Meyer, Joshua and Morais, Reuben and Olayemi, Samuel and others},
  journal={arXiv preprint arXiv:2406.04904},
  year={2024}
}

@article{betker2023better,
  title={Better speech synthesis through scaling},
  author={Betker, James},
  journal={arXiv preprint arXiv:2305.07243},
  year={2023}
}

@article{chen2024vall,
  title={Vall-e 2: Neural codec language models are human parity zero-shot text to speech synthesizers},
  author={Chen, Sanyuan and Liu, Shujie and Zhou, Long and Liu, Yanqing and Tan, Xu and Li, Jinyu and Zhao, Sheng and Qian, Yao and Wei, Furu},
  journal={arXiv preprint arXiv:2406.05370},
  year={2024}
}

@article{wang2023neural,
  title={Neural codec language models are zero-shot text to speech synthesizers},
  author={Wang, Chengyi and Chen, Sanyuan and Wu, Yu and Zhang, Ziqiang and Zhou, Long and Liu, Shujie and Chen, Zhuo and Liu, Yanqing and Wang, Huaming and Li, Jinyu and others},
  journal={arXiv preprint arXiv:2301.02111},
  year={2023}
}

@article{ju2024naturalspeech,
  title={Naturalspeech 3: Zero-shot speech synthesis with factorized codec and diffusion models},
  author={Ju, Zeqian and Wang, Yuancheng and Shen, Kai and Tan, Xu and Xin, Detai and Yang, Dongchao and Liu, Yanqing and Leng, Yichong and Song, Kaitao and Tang, Siliang and others},
  journal={arXiv preprint arXiv:2403.03100},
  year={2024}
}

@article{shen2023naturalspeech,
  title={Naturalspeech 2: Latent diffusion models are natural and zero-shot speech and singing synthesizers},
  author={Shen, Kai and Ju, Zeqian and Tan, Xu and Liu, Yanqing and Leng, Yichong and He, Lei and Qin, Tao and Zhao, Sheng and Bian, Jiang},
  journal={arXiv preprint arXiv:2304.09116},
  year={2023}
}

@article{guo2024fireredtts,
  title={Fireredtts: A foundation text-to-speech framework for industry-level generative speech applications},
  author={Guo, Hao-Han and Hu, Yao and Liu, Kun and Shen, Fei-Yu and Tang, Xu and Wu, Yi-Chen and Xie, Feng-Long and Xie, Kun and Xu, Kai-Tuo},
  journal={arXiv preprint arXiv:2409.03283},
  year={2024}
}

@article{anastassiou2024seed,
  title={Seed-tts: A family of high-quality versatile speech generation models},
  author={Anastassiou, Philip and Chen, Jiawei and Chen, Jitong and Chen, Yuanzhe and Chen, Zhuo and Chen, Ziyi and Cong, Jian and Deng, Lelai and Ding, Chuang and Gao, Lu and others},
  journal={arXiv preprint arXiv:2406.02430},
  year={2024}
}

@article{du2024cosyvoice1,
  title={Cosyvoice: A scalable multilingual zero-shot text-to-speech synthesizer based on supervised semantic tokens},
  author={Du, Zhihao and Chen, Qian and Zhang, Shiliang and Hu, Kai and Lu, Heng and Yang, Yexin and Hu, Hangrui and Zheng, Siqi and Gu, Yue and Ma, Ziyang and others},
  journal={arXiv preprint arXiv:2407.05407},
  year={2024}
}

@article{du2024cosyvoice2,
  title={Cosyvoice 2: Scalable streaming speech synthesis with large language models},
  author={Du, Zhihao and Wang, Yuxuan and Chen, Qian and Shi, Xian and Lv, Xiang and Zhao, Tianyu and Gao, Zhifu and Yang, Yexin and Gao, Changfeng and Wang, Hui and others},
  journal={arXiv preprint arXiv:2412.10117},
  year={2024}
}

@inproceedings{eskimez2024e2,
  title={E2 tts: Embarrassingly easy fully non-autoregressive zero-shot tts},
  author={Eskimez, Sefik Emre and Wang, Xiaofei and Thakker, Manthan and Li, Canrun and Tsai, Chung-Hsien and Xiao, Zhen and Yang, Hemin and Zhu, Zirun and Tang, Min and Tan, Xu and others},
  booktitle={Proc. SLT},
  year={2024},
}

@article{lee2024ditto,
  title={Ditto-tts: Efficient and scalable zero-shot text-to-speech with diffusion transformer},
  author={Lee, Keon and Kim, Dong Won and Kim, Jaehyeon and Cho, Jaewoong},
  journal={arXiv preprint arXiv:2406.11427},
  year={2024}
}

@article{zhang2023speak,
  title={Speak foreign languages with your own voice: Cross-lingual neural codec language modeling},
  author={Zhang, Ziqiang and Zhou, Long and Wang, Chengyi and Chen, Sanyuan and Wu, Yu and Liu, Shujie and Chen, Zhuo and Liu, Yanqing and Wang, Huaming and Li, Jinyu and others},
  journal={arXiv preprint arXiv:2303.03926},
  year={2023}
}

@article{jiang2025megatts,
  title={MegaTTS 3: Sparse Alignment Enhanced Latent Diffusion Transformer for Zero-Shot Speech Synthesis},
  author={Jiang, Ziyue and Ren, Yi and Li, Ruiqi and Ji, Shengpeng and Zhang, Boyang and Ye, Zhenhui and Zhang, Chen and Jionghao, Bai and Yang, Xiaoda and Zuo, Jialong and others},
  journal={arXiv preprint arXiv:2502.18924},
  year={2025}
}

@article{lajszczak2024base,
  title={Base tts: Lessons from building a billion-parameter text-to-speech model on 100k hours of data},
  author={{\L}ajszczak, Mateusz and C{\'a}mbara, Guillermo and Li, Yang and Beyhan, Fatih and Van Korlaar, Arent and Yang, Fan and Joly, Arnaud and Mart{\'\i}n-Cortinas, {\'A}lvaro and Abbas, Ammar and Michalski, Adam and others},
  journal={arXiv preprint arXiv:2402.08093},
  year={2024}
}

@article{kharitonov2023speak,
  title={Speak, read and prompt: High-fidelity text-to-speech with minimal supervision},
  author={Kharitonov, Eugene and Vincent, Damien and Borsos, Zal{\'a}n and Marinier, Rapha{\"e}l and Girgin, Sertan and Pietquin, Olivier and Sharifi, Matt and Tagliasacchi, Marco and Zeghidour, Neil},
  journal={In journal TACL},
  year={2023},
}

@article{yang2024simplespeech1,
  title={Simplespeech: Towards simple and efficient text-to-speech with scalar latent transformer diffusion models},
  author={Yang, Dongchao and Wang, Dingdong and Guo, Haohan and Chen, Xueyuan and Wu, Xixin and Meng, Helen},
  journal={arXiv preprint arXiv:2406.02328},
  year={2024}
}

@article{yang2024simplespeech2,
  title={Simplespeech 2: Towards simple and efficient text-to-speech with flow-based scalar latent transformer diffusion models},
  author={Yang, Dongchao and Huang, Rongjie and Wang, Yuanyuan and Guo, Haohan and Chong, Dading and Liu, Songxiang and Wu, Xixin and Meng, Helen},
  journal={arXiv preprint arXiv:2408.13893},
  year={2024}
}

@inproceedings{du2024unicats,
  title={Unicats: A unified context-aware text-to-speech framework with contextual vq-diffusion and vocoding},
  author={Du, Chenpeng and Guo, Yiwei and Shen, Feiyu and Liu, Zhijun and Liang, Zheng and Chen, Xie and Wang, Shuai and Zhang, Hui and Yu, Kai},
  booktitle={Proc. AAAI},
  year={2024}
}

@inproceedings{song2025ella,
  title={Ella-v: Stable neural codec language modeling with alignment-guided sequence reordering},
  author={Song, Yakun and Chen, Zhuo and Wang, Xiaofei and Ma, Ziyang and Chen, Xie},
  booktitle={Proc. AAAI},
  year={2025}
}

@article{han2024vall,
  title={VALL-E R: Robust and efficient zero-shot text-to-speech synthesis via monotonic alignment},
  author={Han, Bing and Zhou, Long and Liu, Shujie and Chen, Sanyuan and Meng, Lingwei and Qian, Yanming and Liu, Yanqing and Zhao, Sheng and Li, Jinyu and Wei, Furu},
  journal={arXiv preprint arXiv:2406.07855},
  year={2024}
}

@article{yang2025pseudo,
  title={Pseudo-Autoregressive Neural Codec Language Models for Efficient Zero-Shot Text-to-Speech Synthesis},
  author={Yang, Yifan and Liu, Shujie and Li, Jinyu and Hu, Yuxuan and Wu, Haibin and Wang, Hui and Yu, Jianwei and Meng, Lingwei and Sun, Haiyang and Liu, Yanqing and others},
  journal={arXiv preprint arXiv:2504.10352},
  year={2025}
}

@article{xin2024rall,
  title={Rall-e: Robust codec language modeling with chain-of-thought prompting for text-to-speech synthesis},
  author={Xin, Detai and Tan, Xu and Shen, Kai and Ju, Zeqian and Yang, Dongchao and Wang, Yuancheng and Takamichi, Shinnosuke and Saruwatari, Hiroshi and Liu, Shujie and Li, Jinyu and others},
  journal={arXiv preprint arXiv:2404.03204},
  year={2024}
}

@article{ardila2019common,
  title={Common voice: A massively-multilingual speech corpus},
  author={Ardila, Rosana and Branson, Megan and Davis, Kelly and Henretty, Michael and Kohler, Michael and Meyer, Josh and Morais, Reuben and Saunders, Lindsay and Tyers, Francis M and Weber, Gregor},
  journal={arXiv preprint arXiv:1912.06670},
  year={2019}
}

@inproceedings{guo2021didispeech,
  title={Didispeech: A large scale mandarin speech corpus},
  author={Guo, Tingwei and Wen, Cheng and Jiang, Dongwei and Luo, Ne and Zhang, Ruixiong and Zhao, Shuaijiang and Li, Wubo and Gong, Cheng and Zou, Wei and Han, Kun and others},
  booktitle={Proc. ICASSP},
  year={2021},
}

@misc{Iida2021Kokoro,
  author       = {Katsuya Iida},
  title        = {Kokoro Speech Dataset},
  year         = {2021},
  howpublished = {\url{https://github.com/kaiidams/Kokoro-Speech-Dataset}},
  url          = {https://github.com/kaiidams/Kokoro-Speech-Dataset}
}

@article{gao2023funasr,
  title={Funasr: A fundamental end-to-end speech recognition toolkit},
  author={Gao, Zhifu and Li, Zerui and Wang, Jiaming and Luo, Haoneng and Shi, Xian and Chen, Mengzhe and Li, Yabin and Zuo, Lingyun and Du, Zhihao and Xiao, Zhangyu and others},
  journal={arXiv preprint arXiv:2305.11013},
  year={2023}
}

@article{chen2022wavlm,
  title={Wavlm: Large-scale self-supervised pre-training for full stack speech processing},
  author={Chen, Sanyuan and Wang, Chengyi and Chen, Zhengyang and Wu, Yu and Liu, Shujie and Chen, Zhuo and Li, Jinyu and Kanda, Naoyuki and Yoshioka, Takuya and Xiao, Xiong and others},
  journal={In Journal JSTSP},
  year={2022},
}

@article{loshchilov2017decoupled,
  title={Decoupled weight decay regularization},
  author={Loshchilov, Ilya and Hutter, Frank},
  journal={arXiv preprint arXiv:1711.05101},
  year={2017}
}

@article{holtzman2019curious,
  title={The curious case of neural text degeneration},
  author={Holtzman, Ari and Buys, Jan and Du, Li and Forbes, Maxwell and Choi, Yejin},
  journal={arXiv preprint arXiv:1904.09751},
  year={2019}
}

@article{zeghidour2021soundstream,
  title={Soundstream: An end-to-end neural audio codec},
  author={Zeghidour, Neil and Luebs, Alejandro and Omran, Ahmed and Skoglund, Jan and Tagliasacchi, Marco},
  journal={In Journal TASLPRO},
  year={2021},
}

@article{zhang2023speechtokenizer,
  title={Speechtokenizer: Unified speech tokenizer for speech large language models},
  author={Zhang, Xin and Zhang, Dong and Li, Shimin and Zhou, Yaqian and Qiu, Xipeng},
  journal={arXiv preprint arXiv:2308.16692},
  year={2023}
}

@book{Eberhard2024,
  editor    = {Eberhard, David M. and Simons, Gary F. and Fennig, Charles D.},
  title     = {Ethnologue: Languages of the World},
  edition   = {Twenty-seventh},
  year      = {2024},
  publisher = {SIL International},
  address   = {Dallas, Texas},
  url       = {http://www.ethnologue.com}
}

@article{team2023gemini,
  title={Gemini: a family of highly capable multimodal models},
  author={Team, Gemini and Anil, Rohan and Borgeaud, Sebastian and Alayrac, Jean-Baptiste and Yu, Jiahui and Soricut, Radu and Schalkwyk, Johan and Dai, Andrew M and Hauth, Anja and Millican, Katie and others},
  journal={arXiv preprint arXiv:2312.11805},
  year={2023}
}

@inproceedings{mcauliffe2017montreal,
  title={Montreal forced aligner: Trainable text-speech alignment using kaldi.},
  author={McAuliffe, Michael and Socolof, Michaela and Mihuc, Sarah and Wagner, Michael and Sonderegger, Morgan},
  booktitle={Proc. Interspeech},
  year={2017}
}

@article{kingma2014adam,
  title={Adam: A method for stochastic optimization},
  author={Kingma, Diederik P and Ba, Jimmy},
  journal={arXiv preprint arXiv:1412.6980},
  year={2014}
}

@inproceedings{kim2024clam,
  title={CLaM-TTS: Improving Neural Codec Language Model for Zero-Shot Text-to-Speech},
  author={Kim, Jaehyeon and Lee, Keon and Chung, Seungjun and Cho, Jaewoong},
  booktitle={Proc. ICLR},
  year={2024}
}

\clearpage
\appendix

\section{Dataset}
\subsection{Training Dataset Statistics}
The training datasets for each language are as shown in Table~\ref{tab:corpora}. For all of them, we remove all YouTube clips.
\label{app:dataset}
\begin{table}[htp]
  \centering
  \footnotesize
  \caption{Speech-corpus statistics used for training
           (\textbf{total: 32\,578 h}).}
  \label{tab:corpora}
  \renewcommand{\arraystretch}{1.2}

  \begin{tabular}{@{}llr@{}}   % l = language, l = datasets, r = hours
    \toprule
    \textbf{Language} & \textbf{Dataset(s)}    & \textbf{Hours} \\
    \midrule
    \multirow{3}{*}{\textbf{English}}
        & LibriTTS-R~\citep{koizumi2023libritts}  & 516 \\
        & GigaSpeech~\citep{chen2021gigaspeech}  &  5 783                        \\
        & MLS~\citep{pratap2020mls}         &   8 235                       \\
    \midrule
    \multirow{3}{*}{\textbf{Chinese}}
        & WenetSpeech4TTS~\citep{ma2024wenetspeech4tts} & 3 282 \\
        & AISHELL-2~\citep{du2018aishell}       &    997                    \\
        & MAGICDATA~\citep{magicdata}       &       707                 \\
    \midrule
    \textbf{Korean}      & KsponSpeech~\citep{bang2020ksponspeech}         &   832 \\
    \midrule
    \textbf{Japanese}    & ReazonSpeech~\citep{yin2023reazonspeech}        & 3 489 \\
    \midrule
    \textbf{Spanish}     &\multirow{7}{*}{\makecell{MLS~\citep{pratap2020mls} \\ CML-TTS~\citep{oliveira2023cml}}}         & 1 191 \\
    \textbf{French}      &         & 1 338 \\
    \textbf{German}      &         & 3 405 \\
    \textbf{Dutch}       &         & 2 147 \\
    \textbf{Italian}     &         &   294 \\
    \textbf{Portuguese}  &         &   223 \\
    \textbf{Polish}      &         &   139 \\
    \midrule
    \textbf{Total}       &                     & \textbf{32 578} \\
    \bottomrule
  \end{tabular}
\end{table}

\subsection{Speech Editing Dataset}
\label{app:editing_dataset}
To create a comprehensive evaluation set for speech editing, we began by selecting a subset of samples from the Text-to-Speech (TTS) evaluation datasets described in Section~\ref{sec:experiments_setup}. For each language, 100-300 original text samples were chosen.

Unlike RealEdit~\citep{peng2024voicecraft}, which relies on manual, sentence-by-sentence human annotation and modification, a process that limits its scalability across many languages, we employed the powerful multilingual capabilities of the Gemini language model~\citep{team2023gemini} to systematically introduce textual modifications to the original sentences. The goal was to generate edited versions that reflect common editing scenarios. To achieve this, Gemini was instructed to perform exactly one of the following specified operations on each original sentence:
\begin{itemize}
    \item \textbf{Insertion:} Adding a sequence of new words into the original sentence.
    \item \textbf{Deletion:} Removing a sequence of words from the original sentence.
    \item \textbf{Substitution:} Replacing a sequence of words in the original sentence with a new sequence of words.
\end{itemize}
To ensure diversity in the complexity and scope of edits, the length of the modified segments was varied. Specifically, all edits involved at least two contiguous words. The modifications ranged from short (2--3 words), to medium (4--6 words), and occasionally longer spans (7--10 words). We show examples in Table~\ref{tab:examples}.

% \begin{figure*}
%     \label{tab:examples}
%     \centering
%     \caption{Examples of the multilngual speech editing dataset.}
%     \includegraphics[width=1\linewidth]{figs/examples.pdf}
% \end{figure*}

\begin{table*}[htp]
    \centering
    \caption{Examples of the multilingual speech editing dataset.}
    \label{tab:examples}
    \includegraphics[width=1\linewidth]{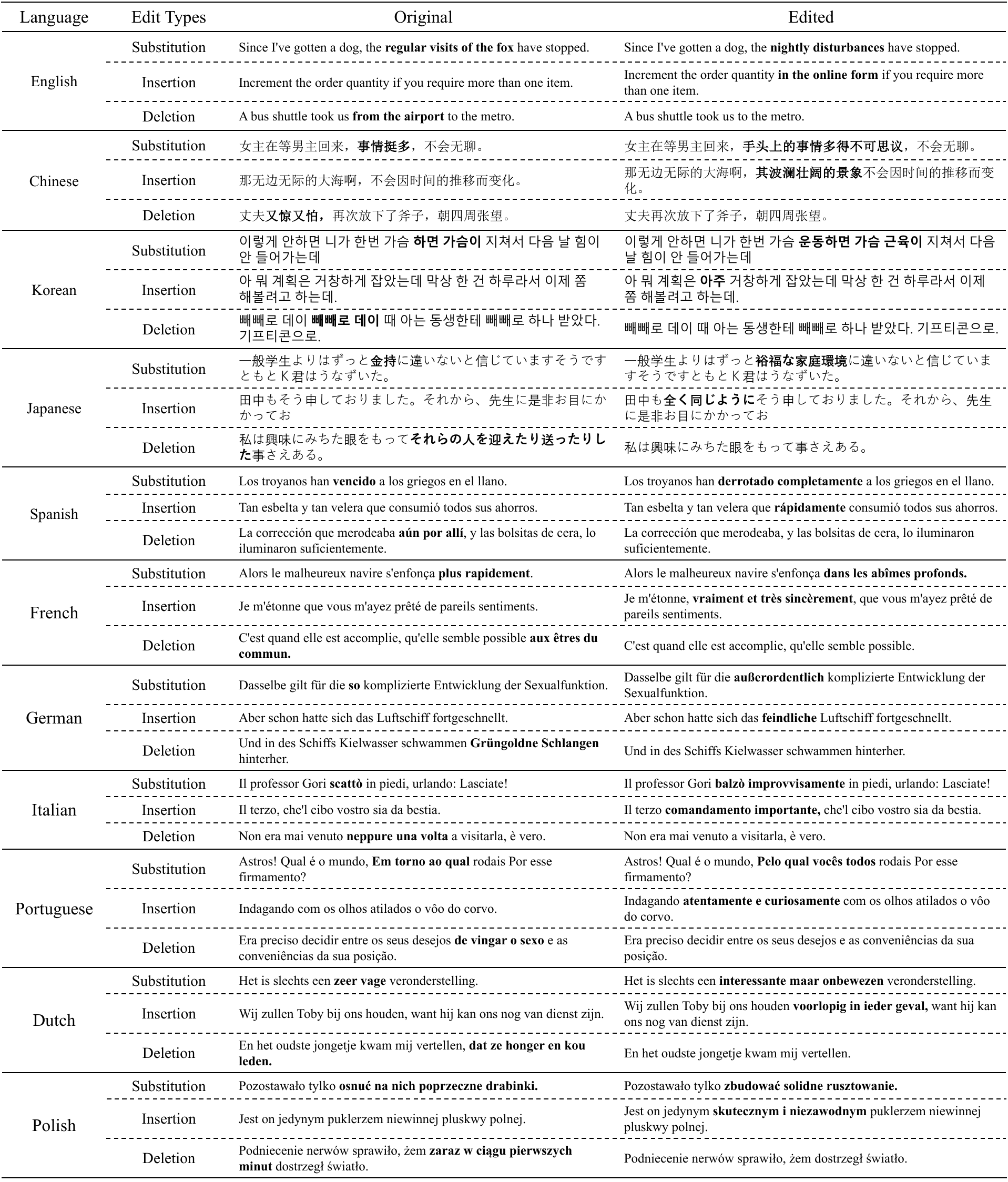}
\end{table*}
\section{Implementational Details}
\label{sec:implement}

\subsection{Encodec Model}\label{app:encodec}
The Encodec model we employ operates with a stride of 320 samples, corresponding to a codec frame rate of 50 Hz when processing audio recorded at 16 kHz. Its encoder begins with a base channel dimension of 64, which doubles at each of the five successive convolutional layers. Following~\citep{defossez2022high}, we utilize the open-source audiocraft repository\footnote{\url{https://github.com/facebookresearch/audiocraft/blob/main/docs/ENCODEC.md}} for training. Specifically, we sample one-second speech segments from the multilingual dataset (shown in Table~\ref{tab:corpora}) and train for 200 epochs with a batch size of 832. Optimization is performed using the Adam algorithm~\citep{kingma2014adam} with a base learning rate of \text{5e-5}.

% \begin{table*}[htbp]
% \centering
% \caption{Comparison of Encodec Models, en/zh/avg}
% \label{tab:encodec_comparison}
% \resizebox{\textwidth}{!}{%
% \begin{tabular}{ccccccccc}
% \toprule
% \textbf{Models} & \textbf{Sampling Rate} & \textbf{Frame rate (Hz)} & \textbf{\#Codebooks} & \textbf{Bandwidth} & \textbf{PESQ $\uparrow$} & \textbf{STOI $\uparrow$} & \textbf{MSTFT $\downarrow$} & \textbf{MCD $\downarrow$} \\
% \midrule
% \multirow{3}{*}{Encodec~\citep{defossez2022high}} & 24000 & 75 & 8  & 6  & 2.717/2.586/2.652 & 0.935/0.930/0.933 & 2.172/1.789/1.981 & 3.019/2.646/2.833 \\
%  & 24000 & 75 & 16 & 12 & 3.329/3.212/3.270 & 0.961/0.957/0.959 & 2.071/1.698/1.884 & 2.391/2.050/2.220 \\
%  & 24000 & 75 & 32 & 24 & 3.651/3.598/3.624 & 0.973/0.968/0.971 & 2.015/1.655/1.835 & 2.016/1.708/1.862 \\\hline
% \multirow{2}{*}{VoiceCraft~\citep{peng2024voicecraft}}      & 16000 & 50 & 4  & 2  & 2.690/2.690/2.690 & 0.933/0.929/0.931 & 1.881/1.662/1.771 & 3.189/2.697/2.943 \\
% & 16000 & 50 & 8  & 4  & 3.085/3.146/3.116 & 0.946/0.943/0.945 & 1.799/1.572/1.686 & 2.840/2.390/2.616 \\\hline
% \multirow{2}{*}{VoiceCraft-X} & 16000 & 50 & 4  & 2  & 2.696/2.774/2.735 & 0.932/0.932/0.932 & 1.874/1.674/1.774 & 3.201/2.688/2.945 \\
%  & 16000 & 50 & 8  & 4  & 3.128/3.225/3.177 & 0.949/0.948/0.948 & 1.811/1.588/1.699 & 2.767/2.316/2.541 \\
% \bottomrule
% \end{tabular}
% }
% \end{table*}

\subsection{Qwen3 Base Model}
\label{app:qwen3_details}
The Qwen3-0.6B-Base model\footnote{\url{https://huggingface.co/Qwen/Qwen3-0.6B-Base}}, foundational to VoiceCraft-X, is a causal language model with 0.6 billion total parameters, of which 0.44 billion are non-embedding parameters. It features 28 Transformer layers, a hidden dimension of 1024, and a feed-forward network (FFN) dimension of 3072, along with 16 attention heads. The model employs Grouped-Query Attention (16 query heads and 8 key/value heads) and supports a context length of 32,768 tokens. A key factor in its suitability for VoiceCraft-X's multilingual requirements is its pre-training on 36 trillion tokens across 119 languages. This pre-training utilized a diverse, high-quality data mix that included multilingual texts, books, and synthetic data. Furthermore, the model incorporates architectural refinements such as \textit{qk layernorm} and benefits from a three-stage pre-training process designed for robust long-context handling.

\subsection{Loss Design}
\label{app:loss_design}
VoiceCraft-X is trained as an autoregressive model to predict a sequence of neural codec tokens. Given the input context, which includes text tokens, speaker embeddings, and potentially prefix/suffix audio tokens, the model predicts the target audio tokens one by one. The overall training objective is a weighted cross-entropy loss, designed to enhance learning efficiency and focus on the crucial aspects of the speech generation task.

Let the sequence of all ground truth speech tokens (encompassing prefix, suffix, and middle segments, and structured according to the delay pattern described in Section~\ref{sec:delay}) be denoted by $Z = (z_1, z_2, \dots, z_N)$, where $N$ is the total number of tokens in the flattened sequence. Each token $z_i$ in this sequence corresponds to an original codec token $Y_{t_i, k_i}$ from timestep $t_i$ and the $k_i$-th codebook of the EnCodec output (where $K=4$ is the total number of codebooks). The model predicts the probability distribution for each token $\hat{z}_i$ conditioned on previous tokens and the input context.

The total loss $\mathcal{L}$ is a sum of individual cross-entropy losses for each token, with two layers of weighting:
\begin{enumerate}
    \item \textbf{Codebook Weighting}: As mentioned in Section~\ref{sec:experiments_setup}, each of the $K=4$ parallel codebooks contributes differently to the overall perceptual quality. We assign weights $\boldsymbol{\alpha} = (\alpha_1, \alpha_2, \alpha_3, \alpha_4) = (1.0, 0.8, 0.6, 0.4)$ to the tokens from codebook 1 to 4, respectively. So, for a token $z_i$ corresponding to $Y_{t_i, k_i}$, its codebook weight is $\alpha_{k_i}$.

    \item \textbf{Segment Weighting}: While the model is trained to predict tokens for all three segments (prefix, middle, and suffix) to improve training efficacy and contextual understanding, the primary goal is the accurate generation of the "middle" (target) segment. To reflect this, we introduce segment-specific weights. Tokens belonging to the "prefix" and "suffix" segments are assigned a weight $w_{seg} = 1$. Tokens belonging to the "middle" segment, which is the primary target for generation or editing, are assigned a higher weight $w_{seg} = 3$. Let $w_{seg}(z_i)$ denote the segment weight for token $z_i$.
\end{enumerate}

Combining these, the total loss $\mathcal{L}$ is formulated as:
$$\mathcal{L} = \sum_{i=1}^{N} w_{seg}(z_i) \cdot \alpha_{k_i} \cdot L_{CE}(\hat{z}_i, z_i)$$
where $L_{CE}(\hat{z}_i, z_i)$ is the cross-entropy loss for predicting token $z_i$. This weighted loss function guides the model to prioritize the generation of the target audio segment while still learning from the context provided by the prefix and suffix, and appropriately valuing the contribution of each codebook.

\section{Subjective Evaluation}
\label{app:subjective}
\subsection{Setup}

To compute our subjective evaluation metrics (SMOS and CMOS for TTS, NMOS and IMOS for Speech Editing), for all languages except Chinese, we recruited Amazon Mechanical Turk workers with a minimum approval rate of $98\%$ and at least $1000$ successful HITs. We manually recruited university students for Chinese. We filtered workers by the following countries in Table~\ref{tab:languages_countries} for each of our languages:
\begin{table}[ht]
\centering
\begin{tabular}{@{}lp{0.6\linewidth}@{}}
\toprule
\textbf{Language} & \textbf{Countries} \\
\midrule
English & United States \\
Chinese & China \\
French & Belgium, Canada, France, Luxembourg, Switzerland \\
Italian & Italy \\
Portuguese & Brazil, Portugal \\
Spanish & Argentina, Chile, Colombia, Mexico, Spain, United States \\
\bottomrule
\end{tabular}
\caption{Countries used to filter crowdworkers for each language}
\label{tab:languages_countries}
\end{table}

\begin{figure*}
    \centering
    \includegraphics[width=0.9\linewidth]{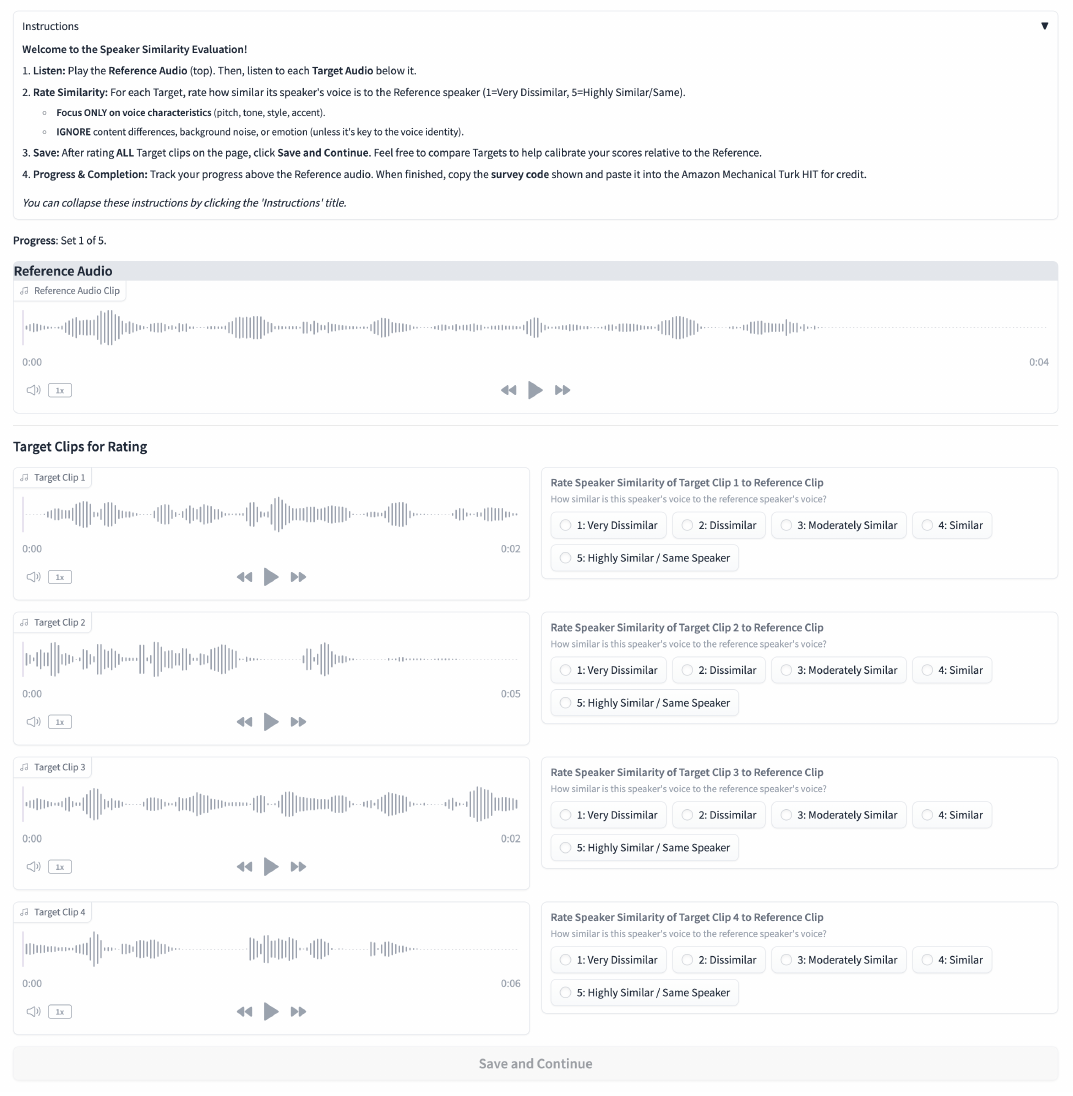}
    \caption{SMOS Annotation UI}
    \label{fig:smos_ui}
\end{figure*}

\begin{figure*}
    \centering
    \includegraphics[width=0.9\linewidth]{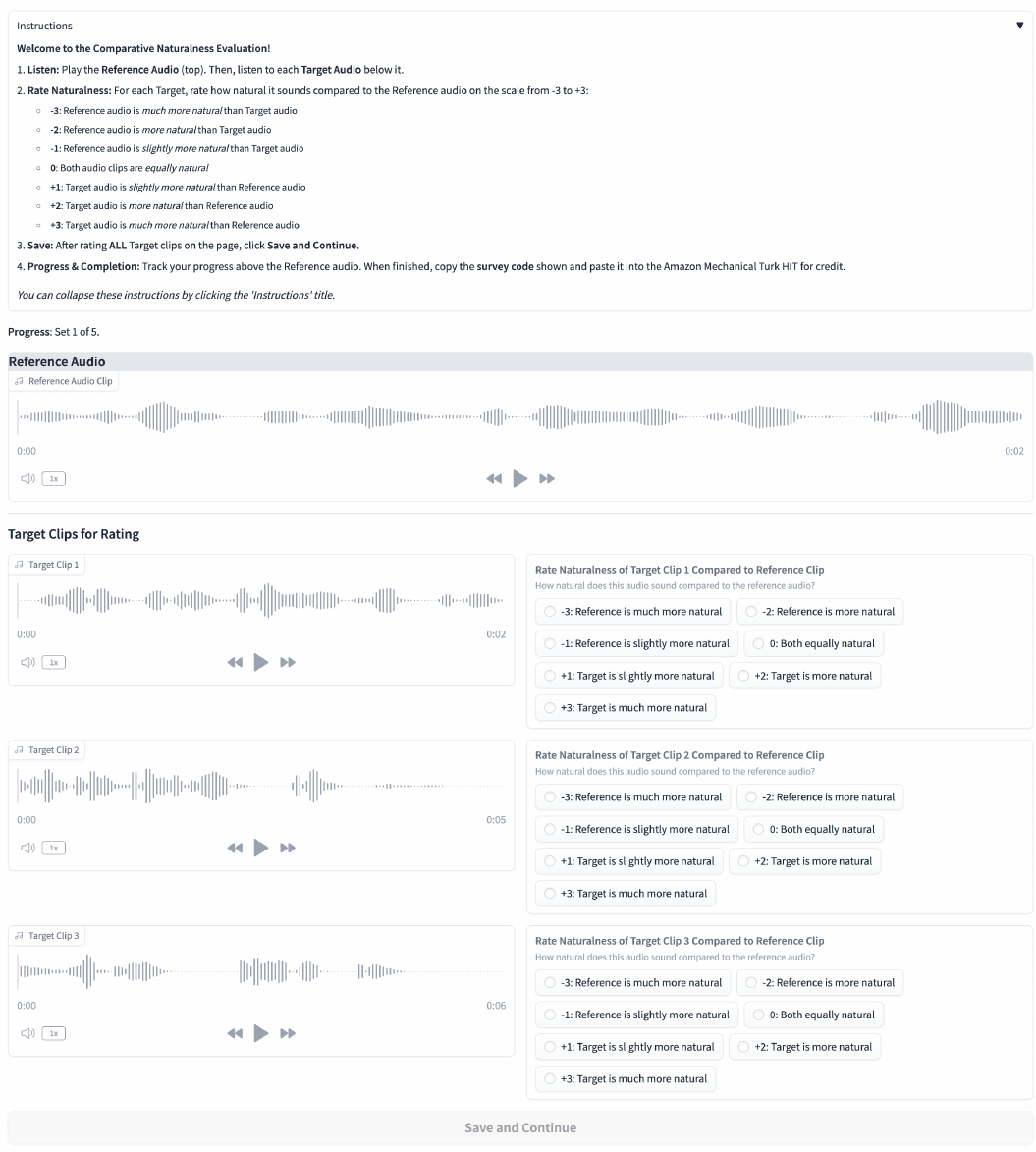}
    \caption{CMOS Annotation UI}
    \label{fig:cmos_ui}
\end{figure*}

\begin{figure*}
    \centering
    \includegraphics[width=0.9\linewidth]{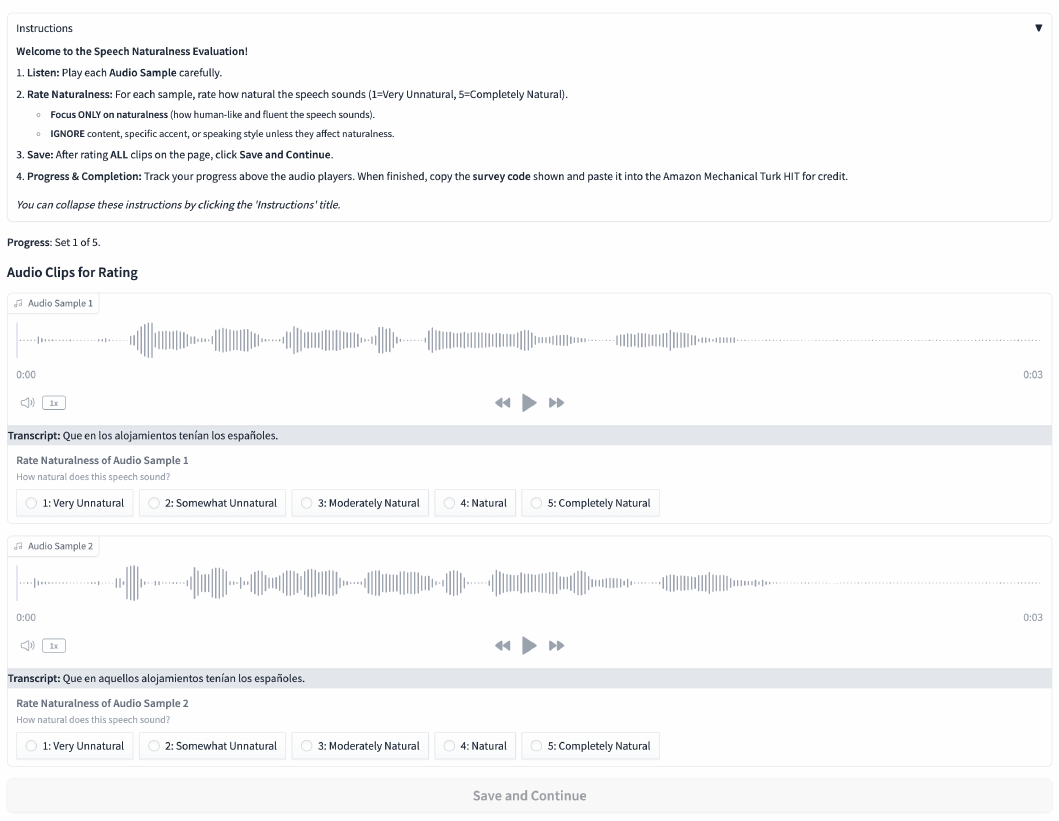}
    \caption{NMOS Annotation UI}
    \label{fig:nmos_ui}
\end{figure*}

\begin{figure*}
    \centering
    \includegraphics[width=0.9\linewidth]{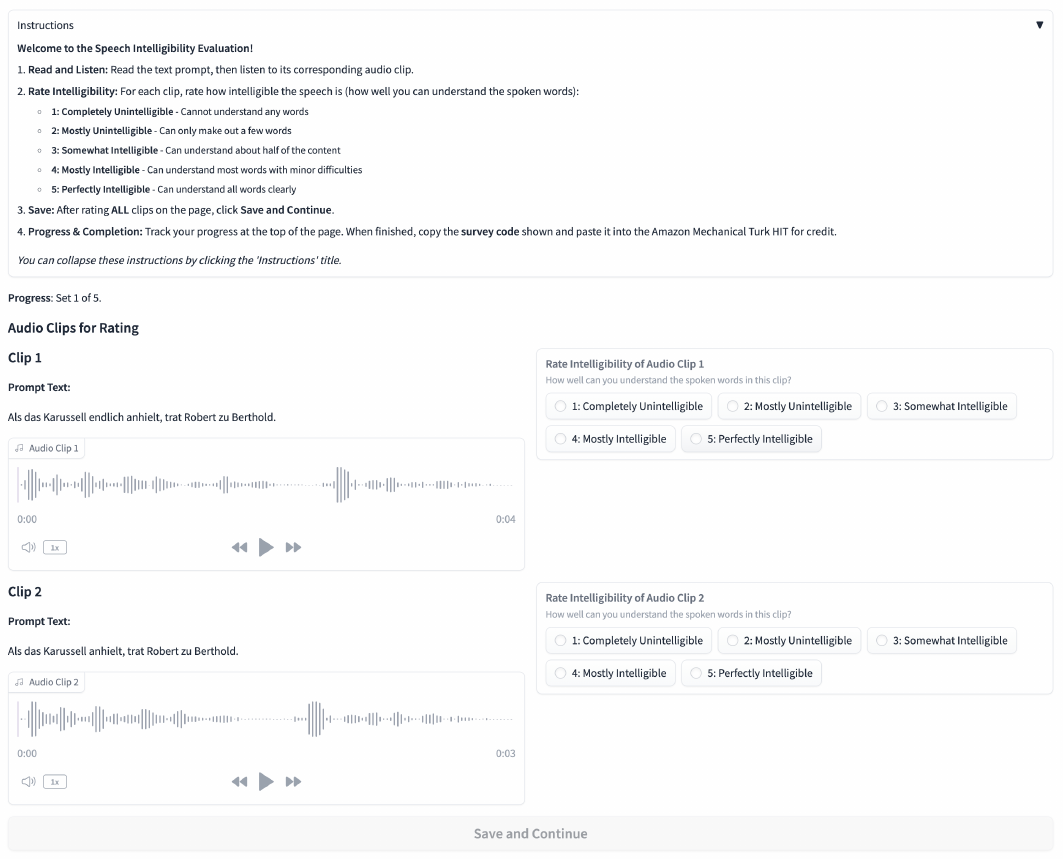}
    \caption{IMOS Annotation UI}
    \label{fig:imos_ui}
\end{figure*}

Each sample was annotated by 3 different annotators. We display annotation UIs for our metrics in Figures~\ref{fig:smos_ui},~\ref{fig:cmos_ui},~\ref{fig:nmos_ui} and~\ref{fig:imos_ui}.

\subsection{Additional Results}
A scarcity of Amazon Mechanical Turk workers for less common languages prevented us from collecting subjective evaluation results for all targeted languages. Consequently, the SMOS results for French, Italian, Portuguese, and Spanish on the Zero-Shot TTS task that we were able to gather are detailed in Table~\ref{tab:addition_tts}.

%For speech editing, NMOS and IMOS results for French, Italian, Portuguese, and Spanish are presented in Table~\ref{tab:addition_speech_editing}.

\begin{table}[htbp]
  \centering
  \footnotesize
  \caption{SMOS on Zero-Shot TTS.}
  \label{tab:addition_tts}
  \begin{tabular}{@{}c c c c c@{}}
    \toprule
    \textbf{Model} & {\textbf{French}} & {\textbf{Italian}} & {\textbf{Portuguese}} & {\textbf{Spanish}}\\
    \midrule
    Ground Truth          & 3.07  & 3.57  & 4.15  & 3.42 \\\hline
    XTTS-v1    & 2.07 & 2.00  & 1.63 & 2.83 \\
    XTTS-v2    & 2.23 & 2.75 & 2.48 & 3.22 \\
    VoiceCraft-X & \textbf{3.58} & \textbf{3.30} & \textbf{2.87} & \textbf{3.58} \\
    \bottomrule
  \end{tabular}
\end{table}

\section{Ablations}
\subsection{Reordering Mechanism}
\begin{table}[H]
\centering
\small
\caption{Impact of token reordering in a low-resource scenario. Models were trained from scratch: one on English (585h LibriTTS-R), the other on Chinese (601h WenetSpeech4TTS Premium subset).}
\begin{tabular}{lcccc}
\toprule
& \multicolumn{2}{c}{English} & \multicolumn{2}{c}{Chinese}\\
\cmidrule(r){2-3} \cmidrule(l){4-5} 
& WER$\downarrow$     & SIM-o$\uparrow$ & CER$\downarrow$ & SIM-o$\uparrow$ \\
\midrule
w/o Reordering & 104.02 & 0.31 & 262.25 & 0.29 \\
w/ Reordering  & \textbf{11.60} & \textbf{0.32} & \textbf{19.25} & \textbf{0.46} \\ 
\bottomrule
\end{tabular}
\label{tab:ordering}
\end{table}

For this ablation study, considering the low-resource nature of most languages, we used LibriTTS-R~\citep{koizumi2023libritts} and the WenetSpeech4TTS Premium~\citep{ma2024wenetspeech4tts} subset as training data. LibriTTS-R contains 585 hours of speech, while the WenetSpeech4TTS Premium subset includes 601 hours\footnote{YouTube clips are removed.}. Models were trained for 15 epochs, both with and without the reordering mechanism. The final epoch was then evaluated on the Seed-TTS test set. As can be seen from Table~\ref{tab:ordering}, the model using the reordering mechanism shows significant performance improvements across all objective evaluation metrics on both the English and Chinese datasets. Specifically, the WER for English dropped dramatically from 104.02 to 11.60, and the CER for Chinese also decreased sharply from 262.25 to 19.25. Concurrently, the SIM-o scores for both languages also showed noticeable increases, indicating an improvement in the quality and naturalness of the synthesized speech. These results strongly demonstrate that the reordering mechanism is very effective in training under low-resource scenarios.

\begin{table*}[htp]
\centering
\caption{WER and SIM-o of different prompt positions in zero-shot TTS inference on Seed-TTS test-en set.} 
\begin{tabular}{lcc}
\toprule
   & WER     & SIM-o   \\
\midrule
 $\varnothing, \varnothing, T_{prompt}, T_{target},\textit{<SPK>},\varnothing,\textit{<M>},\varnothing,\textit{<M>},A_{prompt},A_{target}$ & \textbf{4.37} & \textbf{0.54} \\
 $T_{prompt}, \varnothing, T_{target},\textit{<SPK>},A_{prompt},\textit{<M>},\varnothing,\textit{<M>},A_{target}$ & 5.68 & 0.53 \\
 $\varnothing, T_{prompt}, T_{target},\textit{<SPK>},\varnothing,\textit{<M>},A_{prompt},\textit{<M>},A_{target}$ & 6.32 &	\textbf{0.54}  \\
\bottomrule
\end{tabular}
\label{tab:position}
\end{table*}

\subsection{Position of Prompt in Zero-Shot TTS Inference}
The token reordering mechanism, integral to our training methodology, introduces flexibility in how prompts are structured during zero-shot Text-to-Speech (TTS) inference. To determine the optimal placement, we evaluated several configurations for incorporating the prompt text ($T_{prompt}$) and prompt audio ($A_{prompt}$) into the input sequence. These configurations are detailed in Table~\ref{tab:position}.

Our evaluation, based on WER and SIM-o, revealed that placing the prompt at the beginning of the "middle" segment yields the most favorable overall performance. Specifically, structuring the input such that the prompt text precedes the target text within the middle text segment (i.e., $T_P=\varnothing, T_S=\varnothing, T_M = (T_{prompt}, T_{target})$, with $A_{prompt}$ appended after the mask tokens and before where $A_{target}$ would be generated) resulted in a WER of 4.37, which is notably better than the alternative placements.

\section{Code-Switching}
A desirable characteristic of a multilingual Text-to-Speech (TTS) model is its ability to generate code-switched speech—that is, speech that fluidly transitions between languages. Although our model was trained exclusively on monolingual data, meaning code-switched speech is an out-of-distribution phenomenon for it, the model still demonstrated a certain capacity for code-switching without needing additional language identifiers for inputs in different languages.

We also observed that the model tends to perform better when the initial language of the target text matches the language of the prompt. Conversely, if the starting language of the target text differs from the prompt, the model's performance may be significantly worse. We have made code-switched samples available on our demo page.

%To further investigate the model's code-switching abilities, we fine-tuned it on 10 hours of Chinese–English code-switching data from the ASCEND dataset~\citep{lovenia2021ascend}. We discovered that even with this limited amount of fine-tuning, the model showed a noticeable improvement in Chinese-English code-switching %\david{Can we show a table of objective results/metrics to demonstrate this improvement?}. 

\section{Cross-lingual Finetuning Hours on Zero-Shot TTS}
\begin{figure*}[htb]
    \centering
    \includegraphics[trim={15bp} {0bp} {18bp} {30bp}, clip,width=0.9\linewidth]{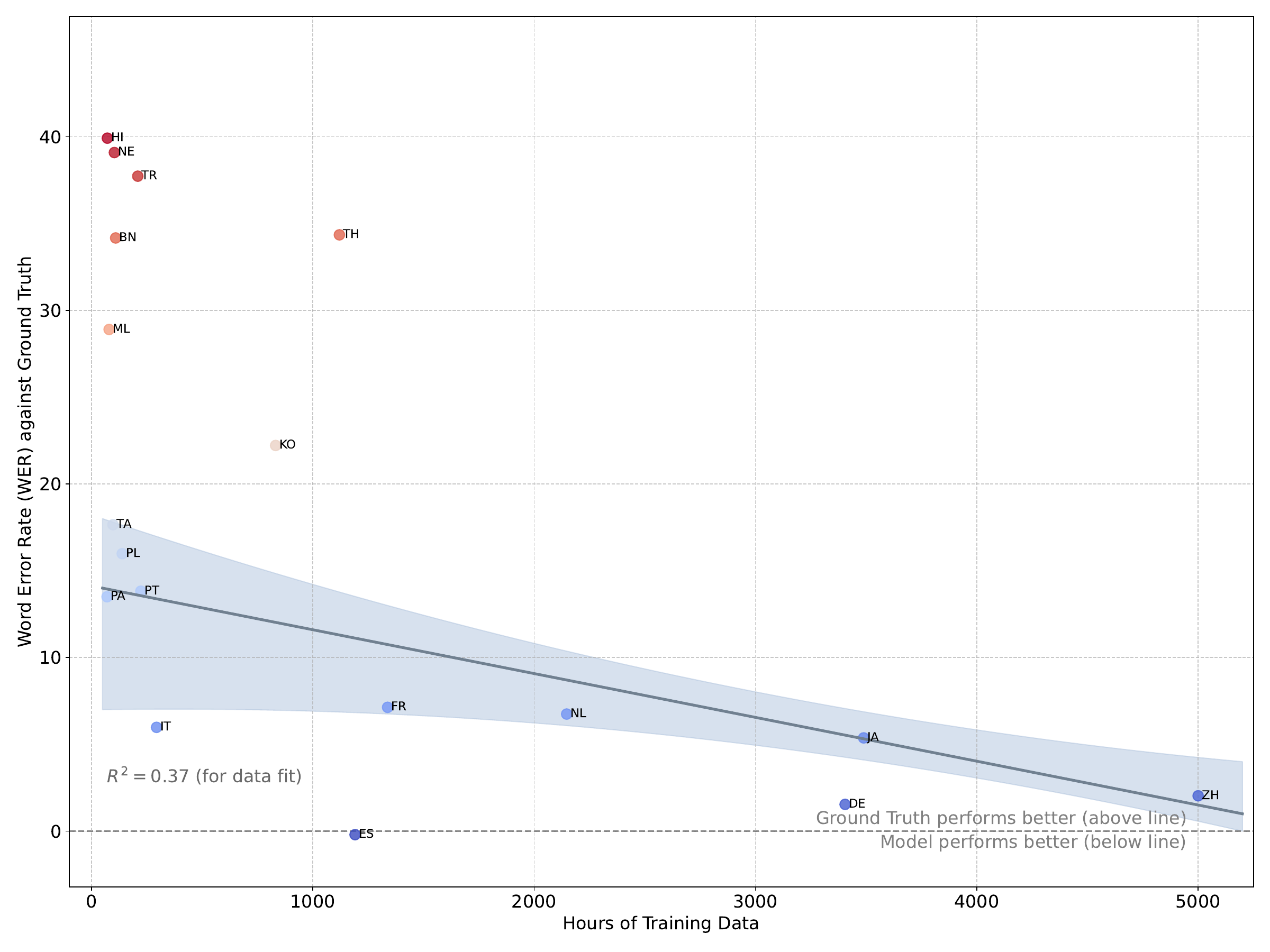}
    \caption{Relationship between per-language fine-tuning data and zero-shot TTS quality. Each point represents a target language, positioned by the number of hours used to fine-tune VoiceCraft-X (x-axis) and the relative Word Error Rate – the difference between Whisper's WER on synthesized audio and its WER on ground-truth audio.}
    \label{fig:language-scale}
\end{figure*}

To further assess VoiceCraft-X's adaptability and the impact of data quantity, we extended fine-tuning experiments across diverse languages. Building on cross-lingual transfer insights (Section~\S\ref{sec:insights_transfer}), we examined the correlation between per-language fine-tuning data volume and zero-shot Text-to-Speech (TTS) quality.

Figure~\ref{fig:language-scale} illustrates these findings, plotting per-language fine-tuning data volume (x-axis) against the relative Word Error Rate (WER) from zero-shot TTS (y-axis). This relative WER, the difference between Whisper's WER on synthesized versus ground-truth audio, offers a normalized measure of intelligibility. The figure generally shows that more fine-tuning data improves pronunciation accuracy, especially for languages sharing similarities with VoiceCraft-X's initial training set. However, this correlation is not universally linear. For languages like Korean and Thai, a moderate data increase (around 1000 hours) did not yield significant WER improvements. This plateauing suggests that for such languages, substantial gains may require much larger or more diverse datasets, or different fine-tuning approaches.


% \section{Polyphone phenomenon}
% investigate some polyphone cases
\end{document}